\documentclass[acmsmall,screen,review=false]{acmart}
\usepackage{color, colortbl}
\usepackage{mathrsfs}
\usepackage{multirow}
\usepackage{graphicx}
\usepackage{subfig}
\usepackage{wrapfig}
\usepackage{geometry}
\usepackage{hyperref}

\definecolor{LightCyan}{rgb}{0.88,1,1}
\newcommand{\blu}[1]{\textcolor{black}{#1}}

\AtBeginDocument{%
  \providecommand\BibTeX{{%
    \normalfont B\kern-0.5em{\scshape i\kern-0.25em b}\kern-0.8em\TeX}}}

\setcopyright{acmcopyright}
\copyrightyear{2018}
\acmYear{2018}
\acmDOI{10.1145/1122445.1122456}

\acmJournal{JACM}
\acmVolume{37}
\acmNumber{4}
\acmArticle{111}
\acmMonth{8}

\begin{document}

\title{Quality change: norm or exception? Measurement, Analysis and Detection of Quality Change in Wikipedia}


\author{Paramita Das}
\email{paramita.das@iitkgp.ac.in}
\affiliation{
  \institution{IIT Kharagpur}
  \city{Kharagpur}
  \country{India-721302}}

\author{Bhanu Prakash Reddy Guda}
\affiliation{
  \institution{Carnegie Mellon University}
  \city{Pittsburgh, PA}
  \country{USA-15213}}

\author{Sasi Bhushan Seelaboyina}
\affiliation{
 \institution{IIT Kharagpur}
 \city{Kharagpur}
 \country{India-721302}}

\author{Soumya Sarkar}
\email{soumya.sarkar@tu-darmstadt.de}
\affiliation{
  \institution{TU Darmstadt}
  \city{Darmstadt}
  \country{Germany-64283}}

\author{Animesh Mukherjee}
\email{animeshm@cse.iitkgp.ac.in}
\affiliation{
  \institution{IIT Kharagpur}
  \city{Kharagpur}
  \country{India-721302}}

\renewcommand{\shortauthors}{P.Das, et al.}
\renewcommand{\shorttitle}{Quality change: norm or exception?}

\begin{abstract}
 Wikipedia has been turned into an immensely popular crowd-sourced encyclopedia for information dissemination on numerous versatile topics in the form of subscription free content. It allows anyone to contribute so that the articles remain comprehensive and updated. For enrichment of content without compromising standards, the Wikipedia community enumerates a detailed set of guidelines, which should be followed. Based on these, articles are categorized into several quality classes by the Wikipedia editors with increasing adherence to guidelines. This quality assessment task by editors is laborious as well as demands platform expertise. As a first objective, in this paper, we study evolution of a Wikipedia article with respect to such quality scales. Our results show novel non-intuitive patterns emerging from this exploration. As a second objective we attempt to develop an automated data driven approach for the detection of the early signals influencing the quality change of articles. We posit this as a \textit{change point detection} problem whereby we represent an article as a time series of consecutive revisions and encode every revision by a set of intuitive features. Finally, various change point detection algorithms are used to efficiently and accurately detect the future change points. We also perform various ablation studies to understand which group of features are most effective in identifying the change points. To the best of our knowledge, this is the first work that rigorously explores English Wikipedia article quality life cycle from the perspective of quality indicators and provides a novel unsupervised page level approach to detect quality switch, which can help in automatic content monitoring in Wikipedia thus contributing significantly to the CSCW community. 
\end{abstract}



\maketitle

\section{Introduction}
Knowledge accumulation and dissemination through peer-production as a form of collective intelligence, working in a decentralized manner without the inclusion of any relational or contractual workforce, has been a predominant source of social good, emerging out of the {\em world wide web}. This also remains one of the prime points of interest of the CSCW community. The peer-production platforms, evolved from the Internet-mediated utilities, has opened up research options in three possible directions of such systems broadly \cite{benkler2015peer}: 1) {\em organization}- the success of the non-hierarchical bureaucracies and the openness in leadership in managing the diverse activities \cite{forte2012coordination,keegan2016analyzing}, 2) {\em motivation}- the inspiration \cite{balestra2017investigating,yu2017predicting} behind the millions of active participants in the absence of personal incentives and 3) {\em quality}- ensuring the rich and reliable products/contents processed though the collective system \cite{warncke2015success}. The first two pillars are deeply rooted with the nature of the rules, guidelines and policies, enforced by the platform. They can be treated as the internal factors, influencing the growth of the system, but the quality profile is the ultimate outcome indeed. The quality can be assessed in various scales and standards that helps in earning the trust of the platform among the consumers. Hence, over the days, researchers are engineering different aspects of quality which in turn is responsible for increase in the viewership of the peer-production system. Likewise the platform's success is directly associated with the quality, and the quality, in turn, is positively correlated to the collaboration patterns within the system. The crowds tend to change their contribution with a varying incoming and outgoing numbers in participation, and hence, the quality itself is dynamic and unstable over time. Understanding the temporal evaluation of quality of large scale peer-production system, for example, Wikipedia could present researchers important clues that are responsible for the success of such a system. 

Wikipedia has emerged as the largest multilingual encyclopedia known to mankind spanning more than 300 languages and $\sim 50M$ articles.
Everyday millions of participants comprising both creators and consumers pour into this platform for keeping themselves updated on a plethora of topics. The platform has been benefited from this synergy and has hence grown in terms of volume and veracity over the last decades. Hence we focus on the English Wikipedia, which is the largest wikiproject covering more than $\sim 6M$ articles and an astonishing $\sim 3B$ words\footnote{\url{https://en.wikipedia.org/wiki/Wikipedia:Size\_of\_Wikipedia}} for our study. Wikipedia's bold policy, ``anyone can edit'' draws attention of the crowd irrespective of the socio-cultural/geographical boundaries to collaborate and contribute openly. Many of them, specially the editors who are concerned with the quality standards assign the articles to the existing quality scales but the quality keeps on fluctuating in the strongly moderated review environment by individuals or panels. Wikipedia's openness often leads to information manipulation and vandalism by inexperienced editors and vandals. In addition, with the vast volume of articles in Wikipedia, most of them are not updated periodically, hence becoming inconsistent and incomplete. Both are serious concerns leading to potential degradation of the encyclopedia quality. To overcome this shortcoming, Wikipedia has implemented user-driven approach to assess the quality of articles. According to Wikipedia's guidelines, an article can be evaluated to any of the quality rankings (FA, A, GA, B, C, Start, Stub; ordered in terms of decreasing quality) by the community of editors. Although the manual assessment includes the perfection of quality assignment task, a major problem that often comes up is of inconsistency. For example, about $4,54,697$ articles have remain unassessed till 2020 in the English Wikipedia.

Moreover, the assessment becomes obsolete quickly with the frequent updation of information. Hence, in reality, the quality of an article is not a static attribute; in contrast it follows a temporal trajectory of improvements as well as declines. In the last few years, Wikimedia foundation has started searching for automatic solutions of quality evaluation, designed for Wikipedia specifically. Leveraging diverse collaborative features of this peer-review system and the structural features of the content, several AI-based (state-of-the-art machine learning and deep learning approaches) techniques have been developed to measure the quality of an article at a specific timestamp. The later solution is able to eradicate the time lag of manual effort but introduces noise in measuring the dynamic change. Depending upon the cost of the system, the automatic prediction framework enables the models to be learned on the entire dataset or partially on few sets of articles. Hence, they fail to capture the dynamic changes of every article individually and create prediction errors. We, in this paper, try to fill the gap between the two approaches - manual and automated by incorporating the dynamic and unstable changes in quality.

There has been a large body of CSCW literature that have attempted to perform a deep dive into Wikipedia~\cite{fearnhead2019changepoint,kittur2007he,kittur2008harnessing,kittur2010beyond,arazy2016turbulent,panciera2009wikipedians}. Our work adds the temporal dimension to the present literature
\cite{wohner2009assessing,hu2007measuring,dang2017end} on the article quality and tries to characterize the quality life-cycle in the peer-reviewed system and also to predict the dynamic changes in the quality states from beforehand. Using a subset of the $30k$ articles, sampled randomly from the English Wikipedia articles spanning over more than $14$ years from the creation of the articles, we build a data pipeline to observe the temporal evolution of the article quality.

\subsection{Research questions}
\begin{enumerate}
    \item \blu{\textbf{R1:} How do the Wikipedia articles transition through different quality states over time?} 
    \item \blu{\textbf{R2:} What will be the appropriate way to detect the dynamic change in article quality? What are some of the most intuitive characteristics of an article that helps in forecasting the upcoming changes in its quality?}
\end{enumerate}

\subsection{Our contributions}
\noindent\textbf{Quality change evolution}: We perform rigorous experiments to understand the fine-grained details of the quality change phenomenon which we tabulate along with their implications in Section~\ref{sec:temporal_analysis}. The closest work to our paper is Zhang et. al.~\cite{zhang2020mining}, where they identified three types of quality trajectory, i.e., {stalled, plateaued and sustained} over a dataset of 6000 articles. One of the major drawback of their approach is that the authors did not use the actual quality of the articles by leveraging the Wikipedia dumps which are evaluated by human experts. Instead they used ORES~\cite{halfaker2019ores} to predict the quality labels which can be potentially erroneous because it does not take into account the temporal changes of article attributes. Moreover, with the static framework of calculating quality at a particular timestamp it achieves at best $62\%$ accuracy \cite{dang2017end,guda2020nwqm}. 
We on the other hand have used the explicit qualities by editors and our results are obtained on a comprehensive dataset of $30k$ articles. 

\noindent\textbf{Change point detection}: Apart from designing rigorous experiments to understand the phenomenon of quality life cycle in Wikipedia, we leverage state-of-the-art multivariate change point detection algorithms~\cite{van2020evaluation}, to solve a novel task of quality change point detection. We combine a series of features from editor attributes, article attributes and activity based features which helps us achieve $76\%$ \textit{covering} in the detection of quality change  points (cf Section~\ref{sec:ChangePoint}). Our approach is unsupervised and page level, unlike traditional approaches of quality prediction where features across pages are used as inputs to a machine learning model~\cite{shen2017hybrid} for predicting quality. One of the drawback of non-page level blackbox quality prediction approaches is that they have relatively worse performance. Even the state-of-the-art technique achieves $63.5\%$ accuracy on this task~\cite{guda2020nwqm}. We take a relatively simpler approach where we aim to detect a quality change as a function of a set of very intuitive features. This approach is significantly lightweight compared to explicit quality prediction approaches~\cite{shen2019joint,guda2020nwqm} which train models with millions of parameters. Moreover our prediction model wins over the existing machine-learning based quality prediction framework of Wikimedia foundation, \textit{ORES} (cf Section \ref{sec:ores}) by a significant margin. We aim to use our change point detection approach to keep potential editors updated about the possibility of quality change and generate suitable alerts for them whenever appropriate. 

\subsection{Key observations}
For \textbf{RQ1}, we find various interesting temporal patterns in the quality of the articles. There are a group of articles that undergo continuous improvement in quality finally reaching the highest quality category {\em featured article} or {\em good article}. Another group of articles undergo interspersed stages of quality improvement, sometimes even making sudden jumps from a very low quality category to a very high quality category and vice versa. In many cases, we observe that there are cycles formed, i.e., starting from a quality category the page undergoes a series of quality changes and comes back to the same category that it started from (aka \textit{cyclic switches}). Surprisingly, almost half of the articles in our dataset remain at the same quality class all through their lifetime without undergoing any promotion or demotion. Almost all of these stagnant cases are low quality articles.


For \textbf{RQ2}, \blu{we observe that at an aggregate level, the \textit{content of an article} digested through a set of article attributes are the prime determinant of its quality. Some of these include the \textit{length of the article}, \textit{the number of references in the article}, \textit{the number of images}, \textit{number of links to other articles}, \textit{presence of infobox} etc. However, when we deep dive, we observe that a mix of \textit{organisational attributes} of the article like \textit{the number of revisions on the talk page}, \textit{the mean time elapsed between two revisions on the talk page} plus a special \textit{content attribute} of the article, i.e., \textit{the ease of its readability in terms of the number of difficult words} act as an even better predictor of quality change points. Effectiveness of such nuanced feature combinations was hitherto unknown.} Change point detection algorithms make use of these attributes to by far outperform the state-of-the-art models of end-to-end quality prediction like ORES. 


Our findings shed new light on automating future peer-production systems. We believe that our work could add significant value to the ongoing drive by Wikipedia to promote participatory machine learning~\cite{halfaker2019ores}. All codes, sample data related to the paper are made available\footnote{\url{https://github.com/sasibhushan3/QualityChangeNormOrException_CSCW}} to promote reproducible research.

\section{Related Work}
A series of studies revealed that the dynamics of collaboration and diverse set of features take care of the creation of high quality content. This topic is well studied in case of Wikipedia, where typical organizational constraints (for example, protection level of pages - semi-protected, unprotected etc.), editors' authority (ex-admin, auto-confirmed users etc.), norms and guidelines (ex-NPOV, three revert rules), technical features, such as bots etc. play key roles behind the individual contribution as well as the quality of end product. Longitudinal analysis of Wikipedia articles have resulted in discovering several key phenomenon which could potentially have either positive or negative impact on the collaborative platform~\cite{zhang2017crowd}. 

\subsection{Longitudinal analysis of Wikipedia}
The community of editors~\cite{panciera2009wikipedians} belonging to a specific wikiproject owns a deeper sense of membership thus enhancing the durability of their contribution and finally leading to the overall growth of the Wikipedia as a whole. Some studies~\cite{arazy2015functional, arazy2017and} established the fact that the organizational structures in peer-production systems are not simple; rather different roles performed by the participants follow a career path which, in turn, confirms their stands in the community.~\citep{burke2008mopping} developed a model that could predict potential candidates to be promoted to role of an \textit{administrator}. A large number of works~\citep{kittur2008harnessing,kittur2010beyond,viegas2007talk, kittur2007he,im2018deliberation} reflected on the co-evolution of coordination and conflict in online knowledge production platforms and the factors behind growth of the conflict. They also illustrated coordination mechanisms to mitigate conflict and outlined the different aspects of successful decentralized environments. Further, collaborative measures~\citep{zhang2017crowd} such as centralization, conflict and experience help to understand the crowd sourcing efforts in uplifting the quality of the articles. Moreover, the improvement is bi-directional; collaboration~\cite{klein2015virtuous} helps in improving article quality, and in turn, it enriches the expertise of the editors. The authors in~\cite{kiesel2017spatio} through spatial and temporal analysis of edit history pointed out key duration of the day and week when anomalous edits are highly likely. In~\cite{yang2016did,arazy2016turbulent} the authors showed distinct editor roles though the study of edit histories and demonstrated their impact on article quality. The authors in~\cite{kane2014emergent} showed that content retention and change in Wikipedia is similar to software development life cycles. Self-organisation of editors to create quality content generation in absence of explicit work flow constraint has been explored in~\cite{arazy2020emergent,jurgens2012temporal} through text and graph based approaches.   

Researchers point ``positive motivation'' as an important pillar behind the success of the encyclopedia. Studies~\cite{geiger2013using,arazy2010determinants,balestra2017fun, morgan2018welcome} show different ways to measure the degree of labour-hours of editors and how their active participation determine the quality. In contrast, the conflict and bureaucracy of the participants can undermine the motivation of contributors~\cite{halfaker2011don}, thus depleting the influx of contributors and ultimately declining the quality of the encyclopedia. To stop the decline of editors, researchers are suggesting socialization policies and incentives~\cite{choi2010socialization,lampe2012classroom,narayan2017wikipedia, morgan2013tea, hsieh2013welcome} for the young as well as experienced editors in order to retain them. In this context, the ongoing Wiki education program~\cite{li2020successful} that includes students from different universities as new editors became successful enough and claims that institutionalized socialization works better than new editor retention. Besides the large pool of registered users, a significant number of anonymous users also participate in editing Wikipedia. The ongoing research~\cite{champion2019forensic,forte2017privacy,jackson2018did} about the edit patterns of anonymous users in controlling the quality is a crucial direction. Although Wikipedia bots play an important role in reducing human labours, their activity cannot be ignored in diverse application of automated patrolling tasks - quality monitoring, vandalism detection etc. The studies in~\cite{zheng2019roles,geiger2017operationalizing} shed light on their activities, e.g., collaboration and conflicts among automated software agents.

\subsection{Content quality assessment in Wikipedia}

Automatic article assessment has been explored by both researchers at Wikimedia foundation~\cite{halfaker2019ores} and academia~\cite{bassani2019automatically}. A popular direction of exploration has been domain specific feature engineering \cite{lipka2010identifying,blumenstock2008size} for training supervised models tailored for this task. For example, the authors in~\cite{hu2009predicting} proposed a two-way automation process - i) selecting high quality articles and nominate them for featured article reviews using interaction features among reviewers, ii) deciding the articles for featured label using machine learning based classifiers. Other approaches explored distributional representation as well as sequence models~\cite{dang2016quality,shen2017hybrid,shen2019joint,zhang2018history,raman2020classifying} where the idea has been to overcome the time consuming manual feature engineering approach. A complementary direction of exploration has been put forward by~\cite{li2015automatically,de2015measuring} where correlation between article quality and structural properties of co-editor network and editor article network has been exploited.

\noindent In this paper we perform longitudinal analysis of quality evolution with respect to how the assessment changes over time which bridges the above two orthogonal directions of research.

\subsection{Change point detection}
In data analysis, detection of change points is a crucial phenomenon from which time series are studied in varying aspects of temporal patterns to detect the exact point of change in the data generation process. From its early inception back in the $50s$~\cite{page1955test} to recent days, with an abundance of its applications in various fields including speech processing, bioinformatics, climatology, finance and network traffic data analysis, a number of algorithms have been discovered in theory. Methods for change point detection (CPD) are roughly categorised into \textit{online} \cite{fan2020online} vs. \textit{offline} \cite{truong2020selective}, \textit{univariate} vs. \textit{multivariate} \cite{lavielle2006detection}, and \textit{model-based} vs. \textit{nonparametric} \cite{sharma2016trend, haynes2017computationally}. 
\blu{Online methods can be implemented in real-time setting in which algorithms run concurrently with the process being monitored and aim to detect the change point as soon as possible after it occurs. In the online setup, the algorithms need to inspect a batch of data samples (say $\epsilon$ data points which can be different in different methods) to be able to determine the change points between the old and the new data points. In contrast, offline algorithms consider the entire time series at once and detect the change points in batch mode. Sometimes, the offline algorithms are called \textit{signal segmentation} because segmentation is performed after the entire signal has been collected. In our work, we viewed the quality change of an article as a signal segmentation task and applied different offline methods to choose the best possible segmentation.}

\blu{Further, our data involves more than one variables, i.e, features and hence we have considered the algorithms that are able to detect multiple change points in a multivariate time series settings.}

\blu{In general, change point detection can be performed in either parametric and non-parametric framework. Parametric analysis necessarily assumes that the underlying distributions belong to some known family, e.g., Gaussian distribution etc. When the underlying assumptions of parametric models are largely unknown for the data at hand, non-parametric approaches can be deployed on any stream of continuous random variables without requiring any prior knowledge of their distribution. We have tried both parametric (BinSeg~\cite{scott1974cluster} and PELT~\cite{killick2012optimal}) and non-parametric (ECP~\cite{matteson2014nonparametric}) techniques for performing multiple change point analysis of multivariate observations.}

\section{Background and Data}\label{sec:Dataset}
\subsection{Wikipedia article quality assessment scheme}

Several critics of the peer-production system have questioned about the quality of the content - whether the large-scale collaboration is potentially capable to maintain the quality standards. Despite the variety of challenges, the peer-production systems try to address the quality issues periodically by the manual or automated reviews and Wikipedia follows the same path. Wikipedia provides a hierarchical quality ranking on grounds of topic coverage, organization and technical style of the articles. The different quality classes are FA (featured article), A, GA (good article), B, C, Start, and Stub\footnote{\url{https://en.wikipedia.org/wiki/Wikipedia:Content\_assessment}}. Here FA is placed at the highest rank - articles that are fairly complete and well written. Stub, on the other hand, has the lowest in quality - very little meaningful content with a need for the improvement in the style. The corresponding quality is mentioned in the talk page of an article. The quality assessments are mainly controlled by the members of the wikiprojects or editors, who tag the quality changes in talk pages of an article. For achieving the highest quality (FA or GA), potential articles are nominated by the editors and later reviewed by the team - individuals or panels. A selective number of articles are then listed in the \textit{WP:Featured articles} or \textit{WP:Good articles} lists. \blu{Wikipedia maintains individual lists of articles that have satisfied all the criteria of \textit{featured}\footnote{\url{https://en.wikipedia.org/wiki/Wikipedia:Featured_articles}} or \textit{good}\footnote{\url{https://en.wikipedia.org/wiki/Wikipedia:Good_articles}} articles in the review process. These lists are periodically updated by the community.}
Such maintenance works need to be performed in a timely manner to satisfy the trust issue of viewers. However, auto-patrolling is not normative in Wikipedia so far and therefore a large volume of articles miss the attention of the peer-review community. Our work tries to bridge this gap by first demonstrating the quality evolution trajectories of an article and then auto-generating early alerts for editors in case a quality change is warranted for the article. 
\subsection{Dataset Description}
Wikimedia foundation provides access to the complete revision history of all the articles of different language versions in the form of Wikidumps\footnote{\url{https://dumps.wikimedia.org/}}. For our work, we have downloaded the first 100 English dumps which are stored as 7z archived xml files. These dumps consume $\sim 8TB$ disk space in uncompressed form and consist of $\sim 6m$ English Wikipedia pages. 
Each uncompressed xml file of size $\sim 80GB$ contains a random collection of $\sim 5k$ Wikipedia pages. Because of its periodic updation, the pages have all the revisions from the date of creation to the last version as of June 2019. We parsed each xml file by the mediawiki xml\footnote{\url{https://pypi.org/project/mwxml/}} parser to find out a sample of articles of \textit{main}\footnote{This contains all the encyclopedia articles, the lists, the encyclopedia redirects etc.} namespace as well as the corresponding talk pages in one linear scan. Specifically, let us assume, the parser encounters a main article $P$; we remember it and try to locate \textit{Talk:$P$} in the later scans or vice-versa. However, if the talk page of $P$ is not present in the current xml file or the other downloaded dumps, we ignore the page $P$ in that case, otherwise we include the page in our dataset. While continuing  this process, we keep  track of the number of articles in each of the quality classes: \textbf{FA}, \textbf{GA}, \textbf{A}, \textbf{B}, \textbf{C}, \textbf{Start} and \textbf{Stub} so as to maintain the class proportions. Although the above mentioned process is quite simple, it is able to extract an approximately balanced number of articles from each of the quality classes (see Table \ref{table:table_1}). As the \textbf{FA} and \textbf{A} articles are limited in number, we have not followed the omission approach outlined above for these two classes; instead we have included all of them. Our investigation shows that none of the previous works have published such a dataset containing the entire revisions of the main pages along with the talk pages. We shall place this data comprising $30k$ articles in the public domain at the end of the review process. 

The articles we have collected through the above mentioned process are free from any kind of selection bias except maintaining the ratio of the articles in the individual quality classes. Furthermore, to verify the generalizability, we computed the category distribution of our $30k$ articles and compared the same with the actual one of the entire English Wikipedia. The articles in our dataset consist of 102666 categories among the existing $1m$ categories of Wikipedia. Further, we ranked the frequently occurring categories, in which \textit{living people} and \textit{American films} are the top ones, covering $20\%$ articles in our dataset. If we consider the whole set of English Wikipedia articles these two categories are again at the top making 16\% of the whole data. We computed the Spearman's rank correlation coefficient ($\rho $) for the two rankings (truncated at top five categories constituting respectively 24\% and 17\% of our dataset and the full Wikipedia) and obtained a $\rho$ value as $0.9$. Thus, we show that our dataset and the results we report based on it should be representative sample of the English Wikipedia.

\begin{table}[h!]
    \centering
    \scalebox{1.0}{
    \begin{tabular}{c|c|c|c|c|c|c|c}
    \hline
    \multicolumn{7}{c|}{Class} &
    \multirow{2}{*}{Total} \\
    \cline{1-7} 
    FA & A & GA & B & C & Start & Stub  \\
    \hline
    3536 & 511 & 5780 & 5335 & 4884 & 5459 & 5321 & 30826 \\
    \hline
    \end{tabular}
    }
    \caption{Count of articles in the respective quality classes.}
    \label{table:table_1}
\end{table}

\subsection{Preprocessing}
First, the content of Wikipedia articles and the talk pages in the form of Wiki Markup Language\footnote{\url{https://en.wikipedia.org/wiki/Wikipedia:Wiki_Markup_Language}} (wiki text) are converted into plain English-like text using a standard python text crawler\footnote{\url{https://github.com/attardi/wikiextractor}}. Special tokens are then used to identify meta contents in the page such as \textit{infobox}, \textit{level 1 section headings}, \textit{level 2 section headings}, \textit{interal wikilink}, \textit{external link}, \textit{inline references}, \textit{footnote template}, \textit{quotation template} and the categories using the mediawiki parser\footnote{\url{https://mwparserfromhell.readthedocs.io/en/latest/}}. As we shall see, these will act as important features in the subsequent parts of the work.

\subsection{Merging of quality classes}
According to the hierarchy\footnote{\url{https://en.wikipedia.org/wiki/Wikipedia:Content_assessment}} of quality class division provided by Wikipedia, every class has a related detailed criteria of standards; however, the classification is based on qualitative measures primarily. For example, in the description of the \textbf{C} class, an instruction in the guidelines reads ``\textit{The article should have some references to reliable sources, but may still have significant problems or require substantial cleanup $\dots$}'' with some indefinite quantifiers. So, it is difficult for machine learning models to distinguish the quality classes from the immediate upper/lower classes. Further, there is extreme sparsity of data in some classes which might not allow us to obtain statistically reliable insights. We therefore club the quality classes as noted in Table \ref{tab:table_2}. 
According to our clubbing scheme the four types of quality classes we arrive at are \textbf{FA}, \textbf{AGA}, \textbf{BC}, \textbf{SS} which can be arranged in increasing order of qualities as follows
\begin{equation}
    \textbf{FA} > \textbf{AGA} > \textbf{BC} > \textbf{SS}
    \label{eq_1}
\end{equation}
Note that this new ordering does not violate the predefined hierarchy of quality classes as determined by the Wikipedia community. 

\begin{table}[h]
    \centering
    \begin{tabular}{|c|c|r|}
    \hline
    Old class & New class & Counts\\
    \hline
        FA & \textbf{FA} & 3536\\
        A,GA & \textbf{AGA} & 6291\\
        B,C & \textbf{BC} & 10219\\
        Start,Stub & \textbf{SS} & 10780\\
    \hline
    \end{tabular}
    \vspace{2mm}
    \caption{Merging of quality classes and the counts of articles of the newly defined quality classes.}
    \label{tab:table_2}
\end{table}

\subsection{Basic characteristics of the quality classes}

\blu{In this section we investigate the characteristics of the high-quality and low-quality articles in terms of (i) topics, (ii) user views, (iii) editions/revisions, (iv) number of collaborators as suggested by the reviewer.
Here, we denoted the articles belonging to FA and AGA quality classes as the high-quality and the articles of SS class as the low-quality ones.}

\noindent \blu{\textbf{Topics}: We ranked the frequently occurring categories in which \textit{living people} and \textit{Americans films} are the top ones. We tried to find similar rank in high and low quality articles, in which high quality articles attain the same ranking as that of the whole dataset.} 

\begin{figure}[h!]
     \centering
     \subfloat[User views: FA class. \label{fig:views_fa}]{
         \includegraphics[width=0.28\textwidth]{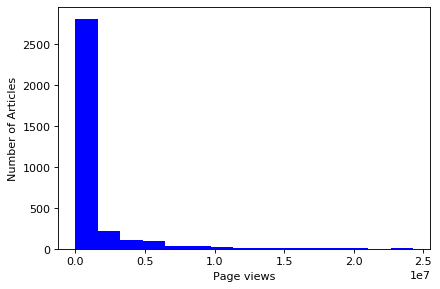}}\qquad
         \subfloat[User views: AGA class.\label{fig:views_aga}]{
              \includegraphics[width=0.28\textwidth]{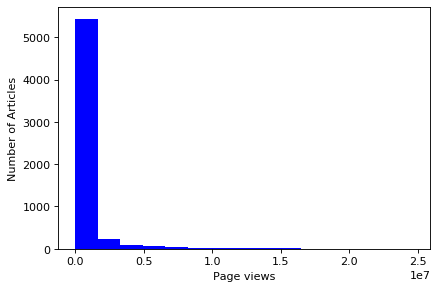}}\qquad
     \subfloat[User views: SS class.\label{fig:views_ss}]{
         \includegraphics[width=0.28\textwidth]{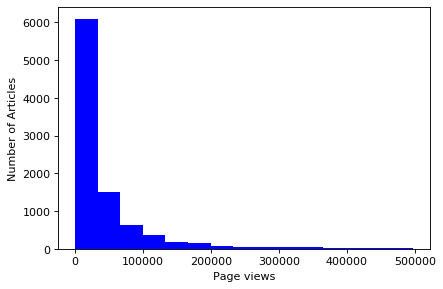}}
        \caption{Histogram showing user views of articles belonging to high (FA, AGA) and low (SS) quality classes.}
        \label{fig:user_views}
\end{figure}

\noindent \blu{\textbf{User views}: We have collected the page views of every article form the Wikimedia API\footnote{\url{https://wikimedia.org/api/rest_v1/\#/}} from the date of availability till June 2019. As expected, the articles belonging to high-quality classes have drawn more user views than the lower ones. The distribution of user views of every class are plotted in the Figure \ref{fig:user_views}.}

\begin{figure}[h!]
     \centering
     \subfloat[Editions: FA class. \label{fig:editions_fa}]{
         \includegraphics[width=0.28\textwidth]{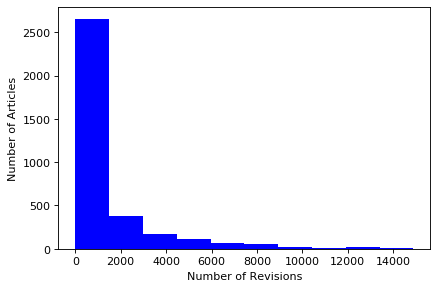}}\qquad
         \subfloat[Editions: AGA class. \label{fig:editions_aga}]{
         \includegraphics[width=0.28\textwidth]{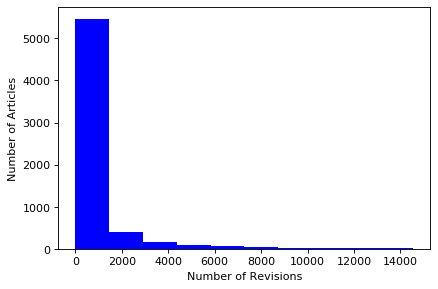}}\qquad
     \subfloat[Editions: SS class.\label{fig:editions_ss}]{
         \includegraphics[width=0.28\textwidth]{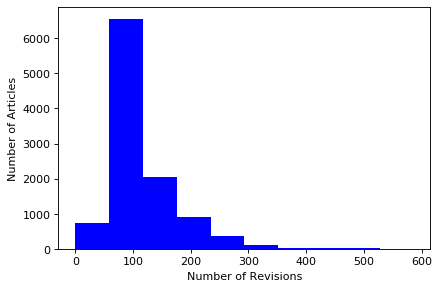}}
        \caption{Histogram showing number of editions/revisions of articles belonging to high (FA, AGA) and low (SS) quality classes.}
        \label{fig:editions}
\end{figure}

\noindent \blu{\textbf{Editions/revisions}: We tried to compare the articles in terms of number of editions/revisions that an article has passed through from the date of the creation. In Figure \ref{fig:editions}, we plotted the distribution of editions for the high and low quality articles in which high quality articles (FA, AGA) have shown larger number of revisions in their life-cycle.}

\begin{figure}[h!]
     \centering
     \subfloat[Collaborators: FA class. \label{fig:collabs_fa}]{
         \includegraphics[width=0.28\textwidth]{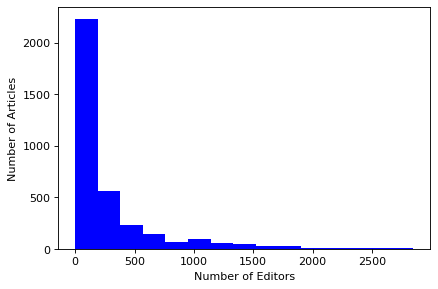}}\qquad
     \subfloat[Collaborators: AGA class.\label{fig:collabs_aga}]{
         \includegraphics[width=0.28\textwidth]{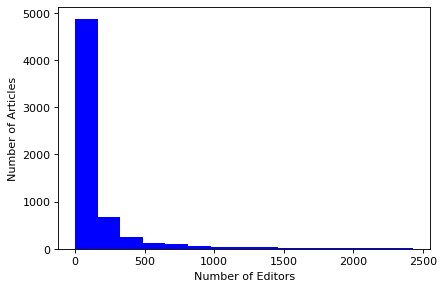}}\qquad
     \subfloat[Collaborators: SS class.\label{fig:collabs_ss}]{
         \includegraphics[width=0.28\textwidth]{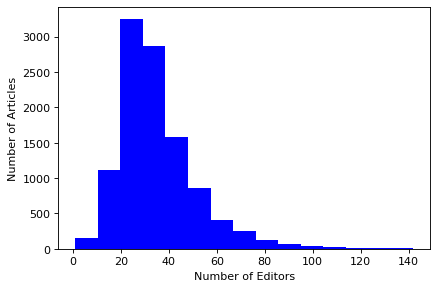}}
        \caption{Histogram showing number of editions/revisions of articles belonging to high (FA, AGA) and low (SS) quality classes.}
        \label{fig:collabs}
\end{figure}

\noindent \blu{\textbf{Number of collaborators}: Similar to the descriptors as mentioned earlier, we have observed the distribution of the number of collaborators for high and low quality articles in Figure \ref{fig:collabs}. The high quality articles have employed larger number of collaborators as compared to the lower quality ones.}

\section{Temporal evolution of article quality} \label{sec:temporal_analysis}
Although measuring Wikipedia article quality is a complex venture, the quality assessment task by the editorial team of typical articles or wikiprojects prioritize further contribution to the articles, influencing editors to revamp continuously. In this section we carry out a detailed analysis of the typical temporal patterns exhibited by the change of qualities of the articles in our dataset.

\subsection{Only promotion}
Let us assume, an article $P$ had the quality \textbf{SS} assigned to it at timestamp $t_{x}$, the quality \textbf{BC} at timestamp $t_{y}$ and finally if the current quality of the article $P$ is \textbf{FA}, where $t_{x} < t_{y}$, we categorize the change of quality as constant \textit{promotion} and the article $P$ is placed in this category. Note that promotions can also skip intermediate classes (e.g., direct \textbf{SS}$\Rightarrow$\textbf{FA}). Following this criteria, we observed that 13249 articles have experienced only promotions in qualities over the revisions and the detailed statistics is noted in Table \ref{tab:table_3}. A page with the sequence of temporal change $\textbf{SS} \Rightarrow \textbf{BC} \Rightarrow \textbf{FA}$ is counted twice in the sub-categories $\textbf{SS} \Rightarrow \textbf{BC}$ and $\textbf{BC} \Rightarrow \textbf{FA}$ individually. 
\noindent
We observe that 42.97\% articles in our dataset undergo one or more promotions. Surprisingly, we found a number of articles falling under the sub-category \textbf{SS} $\Rightarrow$ \textbf{FA} \blu{have the mean time (avg. time) for quality change more than 5 years. Although the standard deviation (SD) reported in this sub-category is slightly higher compared to some other sub-categories (e.g., \textbf{SS} $\Rightarrow$ \textbf{BC}, \textbf{SS} $\Rightarrow$ \textbf{AGA}), it is quite evident that many pages go unnoticed for as long as 5 years or even more resulting in very delayed quality assessment (see the box-plots in Appendix~\ref{appendix:box} for further clarity).} The promotion from \textbf{SS} to \textbf{BC} takes the second highest average time. The fastest category upgradation (in terms of mean and standard deviation of average turn around time) is naturally from \textbf{AGA} $\Rightarrow$ \textbf{FA} which attracts the largest attention of the editors.

\begin{table}[h]
    \centering
    \begin{tabular}{|c|r|r|r|r|}
    \hline
        Type & Number of hops & Count & Avg time (in days) & SD (in days) \\
        \hline
        \rowcolor{LightCyan}
         $\textbf{SS} \Rightarrow \textbf{BC}$ & 1 & 8594 & 1253.70 & 1165.91\\
         \hline
         $\textbf{BC} \Rightarrow \textbf{AGA}$ & 1 & 6144 & 390.58 & 728.73\\
         \hline
         \rowcolor{LightCyan}
         $\textbf{AGA} \Rightarrow \textbf{FA}$ & 1 & 2283 & 294.18 & 476.34 \\
         \hline
         $\textbf{SS} \Rightarrow \textbf{AGA}$ & 2 & 1384 & 1198.25 & 1206.46 \\
         \hline
         $\textbf{BC} \Rightarrow \textbf{FA}$ & 2 & 487 & 535.28 & 881.65 \\
         \hline
         \rowcolor{LightCyan}
         $\textbf{SS} \Rightarrow \textbf{FA}$ & 3 & 97 & 1873.44 & 1380.87\\
         \hline
    \end{tabular}
    \vspace{1mm}
    \caption{Count of articles with only promotion. Highlighted rows show changes that draw special attention.}
    \label{tab:table_3}
\end{table}

\subsection{Only demotion}
We have observed that the temporal sequences of 221 articles in our dataset underwent one or more demotions in quality classes. Likewise \textit{only promotions}, a fall in any number of quality classes have been considered as \textit{only demotion}. Table \ref{tab:table_4} shows overall statistics of this category of temporal change. As in the \textit{only promotion} category, a page with demotions greater than one hop is counted in each of the sub-categories individually.

\noindent
Although only a handful of articles get demoted over the time, such demotions specially for the sub-category \textbf{FA} $\Rightarrow$ \textbf{BC} with an average time gap of quality assessment more than 2 years indicates a surprising exception in article quality changes in Wikipedia. On manual inspection we found that the \textbf{FA} pages in this category have lacked quality contribution for a long time and were hence demoted to lower classes.

\begin{table}[h!]
    \centering
    \begin{tabular}{|c|r|r|r|r|}
    \hline
        Type & Number of hops & Count & Avg time (in days) & SD (in days) \\
        \hline
         $\textbf{BC} \Rightarrow \textbf{SS}$ & 1 & 83 & 542.86 & 655.51\\
         \hline
         $\textbf{AGA} \Rightarrow \textbf{BC}$ & 1 & 105 & 400.33 & 401.32 \\
         \hline
         $\textbf{FA} \Rightarrow \textbf{AGA}$ & 1 & 2 & 469.88 & 294.94 \\
         \hline
         \rowcolor{LightCyan}
         $\textbf{AGA} \Rightarrow \textbf{SS}$ & 2 & 1 & 21.92 & 0 \\
         \hline
         \rowcolor{LightCyan}
         $\textbf{FA} \Rightarrow \textbf{BC}$ & 2 & 32 & 753.54 & 694.44\\
         \hline
         $\textbf{FA} \Rightarrow \textbf{SS}$ & 3 & 0 & 0 & 0\\
         \hline
    \end{tabular}
    \vspace{2mm}
    \caption{Count of articles with only demotion. The highlighted row denotes rare cases of demotion in quality.}
    \label{tab:table_4}
\end{table}

\subsection{Both promotion and demotion}
This set of pages indicate a sequence of promotion and demotion over the revisions in their lifetime. There are 1407 such articles in our dataset.

\subsection{No change in quality}
Surprisingly, there exist a large fraction of articles that have undergone the quality evaluation process only once in their entire life time. Hence, their quality remains intact over the rest of the time frame in our analysis. We include such 15949 pages in this category and majority of these pages come in the quality class \textbf{SS}. The distribution of articles of different quality classes is mentioned in the Table \ref{tab:table_5}. The mean time denotes the average time gap between the creation time of the pages and the time of the first quality assessment. \textbf{FA} pages experienced the highest mean time for the first evaluation. The existence of a large fraction (51.73\%) of the articles in this category clearly depicts that majority of the articles have been overlooked by the current quality assessment framework which is possibly becoming a norm on this platform. Overall this may not be a good sign for the health of the platform.

\begin{table}[h]
    \centering
    \begin{tabular}{|c|r|r|r|}
    \hline
     Quality Class & Count & Mean time (in days) & SD (in days) \\
     \hline
     \textbf{FA} & 341 & 2039.14 & 1750.22 \\
     \hline
     \textbf{AGA} & 375 & 1165.19 & 1610.10\\
     \hline
     \textbf{BC} & 4586 & 935.45 & 1075.24 \\
     \hline
     \textbf{SS} & 10647 & 316.02 & 581.87\\
     \hline
    \end{tabular}
    \vspace{2mm}
    \caption{Count of articles with no change in quality.}
    \label{tab:table_5}
\end{table}

\begin{figure}
    \centering
    \includegraphics[scale=0.35]{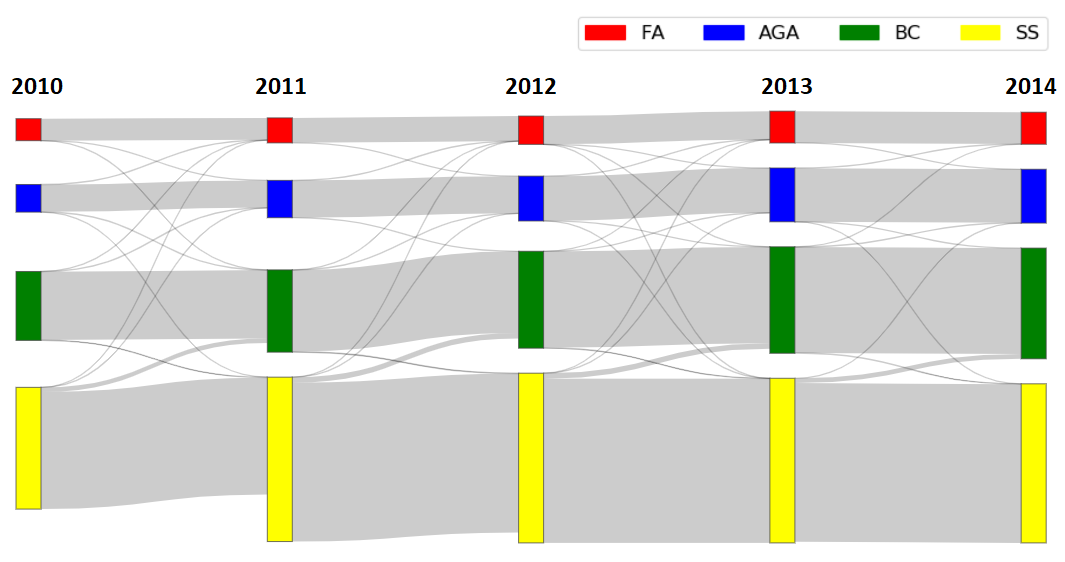}
    \caption{Temporal illustration of intra and inter class quality changes over the years 2010-2014.}
    \label{fig:alluvial_1}
\end{figure}

\begin{figure}
    \centering
    \includegraphics[scale=0.35]{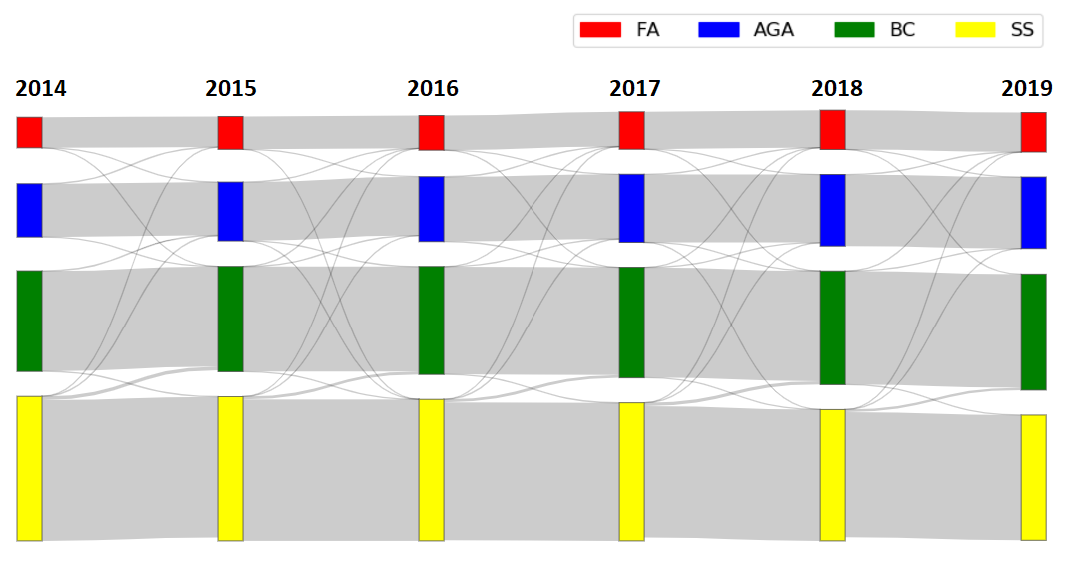}
    \caption{Temporal illustration of intra and inter class quality changes over the years 2014-2019.}
    \label{fig:alluvial_2}
\end{figure}

The \textbf{alluvial diagrams} in Figure \ref{fig:alluvial_1} and \ref{fig:alluvial_2} present an illustration of the temporal change of quality classes in two different time windows: 2010-2014 and 2014-2019. The figures further visually make it evident that more often than not the quality of an article remains fixed over time; only in rare exceptions they exhibit a change.

\subsection{Cyclic switch of qualities}\label{sec:Cyclic}
Let us assume an article $P$ has been assigned to the following quality classes at different timestamps (in an increasing timeline)
\begin{equation}
    \textbf{class1} \Rightarrow class2 \Rightarrow class3 \Rightarrow \textbf{class1}
    \label{eq_2}
\end{equation}
and, recursively,
\begin{equation}
    \textbf{class1} \Rightarrow (classX)^{+} \Rightarrow \textbf{class1}
    \label{eq_3}
\end{equation}
According to equation \ref{eq_2}, the \textit{class1} can be any among the four types of quality classes and \textit{class2}, \textit{class3} can include any permutation of remaining three classes $(P^{3}_{2})$. We have imposed the constraint that the intermediate consecutive class levels can not belong to the same class, for example, \textit{class2} and \textit{class3} cannot be same. We denoted this type of temporal pattern of quality assignment as \textit{cyclic switches} and in equation \ref{eq_2}, the length of the cyclic switch is 4. Such cyclic switches might at times be a possible outcome of editorial conflicts as opposed to organic quality shifts. 

\begin{figure}[h]%
    \centering
        \subfloat[\centering \label{fig:cyclic_1}Distribution of articles, undergoing different number of cyclic switches.]{{\includegraphics[scale=0.45]{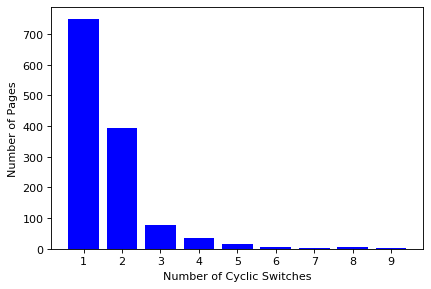}}}%
    \qquad
    \subfloat[\label{fig:cyclic_2}\centering Distribution of length three cyclic switches with turnaround time less than 15 days.]{{\includegraphics[scale= 0.45]{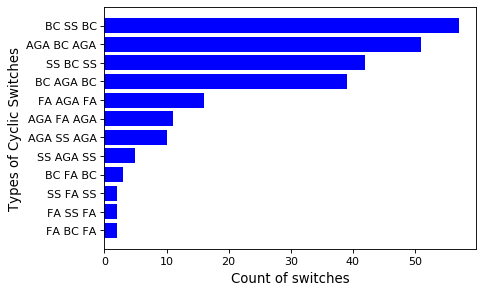}}}%

    \caption{Different distributions of cyclic switches.}
    \label{fig:cyclic_switch}
\end{figure}

\noindent\textbf{Quality switch war?} We found $1286$ articles that exhibited one or more cyclic switches. A majority of the articles have switches of length three ($1258$ articles in all)\footnote{\textcolor{blue}{From the unclubbed to the clubbed version, $\sim90\%$ of the switches are retained.}}. 
The histogram in Figure \ref{fig:cyclic_1} presents the distribution of articles that had undergone different number of cyclic switches. As expected the distribution decays very fast.
\blu{The average turnaround time for cyclic switches in 1286 articles (e.g., length three switches which constitutes the majority) is observed to be $920.89$ days. At the same time, we have observed 180 pages out of the 1286 articles contain cycle switches in which the minimum duration of turnaround time is less than 15 days. We therefore investigate the distribution of these \textit{very rapid} cyclic switches in Figure \ref{fig:cyclic_2}.} In addition, we also saw that wikibots\footnote{\url{https://en.wikipedia.org/wiki/Wikipedia:Bots}} are responsible for cyclic changes in $0.05\%$ cases only compared to the humans. These results together point to the fact that an article experiencing multiple cyclic switches often with short turnaround times in its entire lifespan is an extremely non-trivial behaviour. Likewise edit war this behaviour can be attributed to \textit{quality switch war} that arises possibly due to the continuous conflicts among the editors regarding their perception about the quality of that article. 

\if{0}\begin{figure}[h!]
    \centering
    \includegraphics[scale= 0.5]{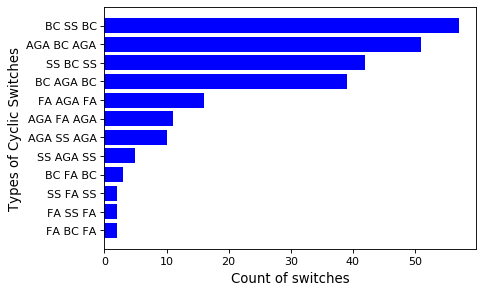}
    \caption{Distribution of length three cyclic switches with turnaround time less than 15 days.}
    \label{fig:cyclic_2}
\end{figure}\fi

\noindent\textbf{Prevalent switches}: Further analysis, tabulated in Table \ref{tab:cyc_tab_1} shows that the count of cyclic switches decreases exponentially with the increase in the length of the switch. We also found that the following cyclic switches <\textbf{BC}, \textbf{SS}, \textbf{BC}>, <\textbf{SS}, \textbf{BC}, \textbf{SS}>, <\textbf{BC}, \textbf{AGA}, \textbf{BC}>, <\textbf{AGA}, \textbf{BC}, \textbf{AGA}> were the top four (in that order) length three switches in terms of occurrence across different articles. This indicates that typically low quality articles are vulnerable to more frequent switches. 

\begin{figure}[h]
    \centering
    \includegraphics[scale=0.6]{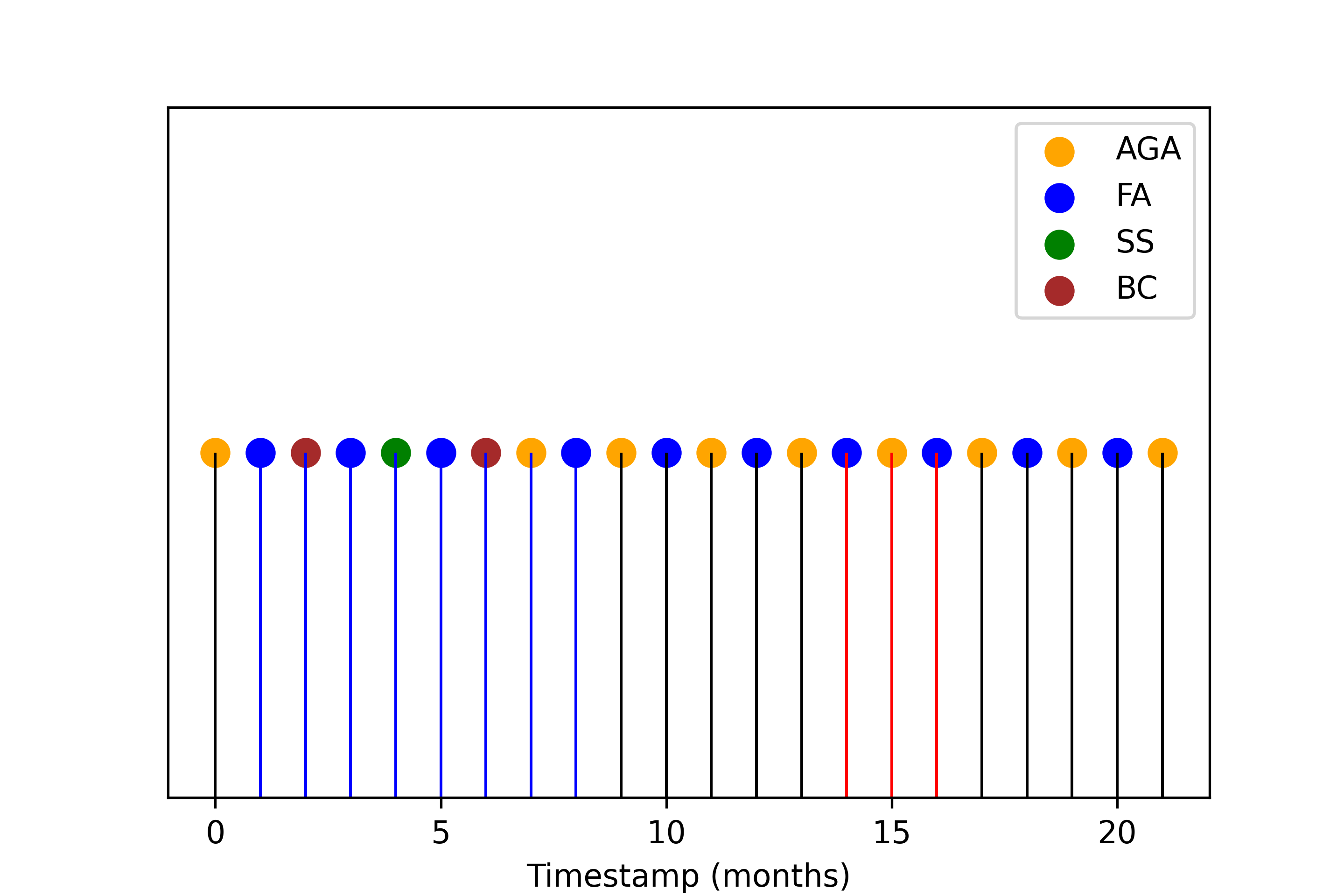}
    \caption{Qualities assigned to an article at different time points. The region plotted in \textcolor{blue}{blue} lines indicate a cyclic switch [FA $\rightarrow$ FA] with larger number of state changes to get back to the initial state compared to the region indicated by the \textcolor{red}{red} lines. The black lines indicate the assigned quality at a particular time but do not correspond to a switching behaviour.}
    \label{fig:cyclic_timestamps}
\end{figure}

\noindent\textbf{Long and short cyclic switches}:
\blu{Based on a closer inspection into the talk pages of some of the articles, we found that cyclic quality switches can be of different lengths. For example, we have observed instances of cyclic switches [FA$\rightarrow$ FA] of varying lengths, i.e., three and eight respectively within a single article (see Figure~\ref{fig:cyclic_timestamps})}.


\begin{table}[h]
    \centering
    \begin{tabular}{|c|r|r|r|r|r|r|r|}
    \hline
    Length & 3 & 4 & 5 & 6 & 7 & 8 & 9  \\
    \hline
    Count & 1993 & 70 & 32 & 4 & 6 & 1 & 1  \\
    \hline
    \end{tabular}
    \vspace{1mm}
    \caption{Count of cyclic switches of varying length of switches.}
    \label{tab:cyc_tab_1}
\end{table}

\section{Detection of Quality Change Points}\label{sec:ChangePoint}
The extensive analysis in the previous sections show that quality assessments of Wikipedia pages experience various hindrances over their lifespan. These range from frequent and unpredictable switches of quality to absolute stagnancy. Therefore, an early alert system that can provide timely signals to the editors suggesting the requirement of quality assessment of an article is very much needed. In this section we attempt to bridge this gap by building an automated unsupervised feature-based approach to discern the change points of article quality as early as possible which would be the first step to developing such an alert system.

\subsection{The quality indicators}
We have carried out a detailed analysis of the typical factors through which article quality evolves over time and based on these have attempted to formulate very intuitive categorical features that could be responsible for such changes. The features that we selected can be grouped into three classes.
\begin{itemize}
    \item Contribution based features (editors' participation attributes) [{\em aka} $G_{c}$]
    \item Activity based features (edit pattern attributes)  [{\em aka} $G_{a}$]
    \item Content based features  (article's attributes) [{\em aka} $G_{p}$]
\end{itemize}

\blu{
For extracting \textit{contribution based features $(G_{c})$}, we have parsed every revision of the articles (an xml file containing all the revisions of an article from the date of the creation) in our dataset and collected the editors' usernames editing the revisions. In case of unregistered editors, the contributor's username is mentioned as anonymous IP address. Similarly, we collected the editors' information from the revision of talk pages of articles individually. The count of the editors, such as distinct registered (unregistered) editors editing the articles and talk pages are used as the features. For extracting \textit{activity based features $(G_{a})$}, the number of revisions an article and its talk page passed through (calculated per month and week basis), mean and variance of time difference between two consecutive revisions of both the article page and talk page at the granularity of months have been calculated. Similar to the features under $(G_{c})$, we have parsed the revisions (i.e., xml files) of article pages and talk pages to collect the activity related features. In case of \textit{content based features $(G_{p})$}, we extracted the content of the latest revision of the articles in every month and computed features of this category. We have used the mediawiki\footnote{\url{https://pypi.org/project/mwparserfromhell/}} parser to parse the wiki text and computed various content based features.} In the following we present an elaborate description of all these features.

\subsubsection{Contribution based features $(G_{c})$}
The features we include in this category describe the involvement of the editors over the revisions of an article. The choice of these features is motivated by the fact that the contribution of editors should have some impact on the quality change of an article. \citeauthor{kittur2008harnessing}~\cite{kittur2008harnessing} examined how the number of editors and their explicit and implicit coordination influence in changing the quality of articles. Also, their research showed that a large pool of editors can play a significant role in quality moderation compared to fewer number of contributors. But in every case, either implicit or explicit coordination is required. With implicit coordination, a core group of editors perform the major work and peripheral editors support them. On the other hand, explicit coordination through the communication on the article talk pages is associated with quality change of an article. We replicated their view on editors' participation in our work. Our hypothesis is that the temporal changes in the number of editors contributing to an article could bear early signs of quality shift. 
\begin{itemize}
    \item \textit{The number of distinct registered editors editing the talk pages} (\textbf{F1}) : Talk pages act like a discussion forum among the editors regarding the content, organization of the specific article and the participation of editors in discussions typically get enhanced close to points of quality change.
    \item \textit{The number of newly added registered editors editing talk pages} (\textbf{F2}) : Similar to the previous feature, we consider the count of the new editors commenting in the talk pages as one of the contribution based features.
    \item \textit{The number of distinct unregistered editors editing the talk pages} (\textbf{F3}) : Once again, our hypothesis is that a large number of unregistered editors (anonymous IPs) participating in the talk page discussions of an article at a point should be indicative of a quality change in the near future.
    \item \textit{The number of distinct registered editors editing the article main page} (\textbf{F4}) : The increase/decrease in the number of editors, editing the main content should impact quality change.
    \item \textit{The number of newly added registered editors editing the main page} (\textbf{F5}) : We use the count of new editors editing an article (suggestive of enhanced attention because of the ongoing review process or a sudden increase in popularity of the respective article) as a quality change indicator.
    \item \textit{The number of unregistered editors editing the article main page} (\textbf{F6}) : A sudden change in the number of unregistered editors (anonymous IPs), editing the main content should be indicative of quality change.

\end{itemize}

We have not considered the editors' experience or their level of engagement in our contribution based features to reduce the noise in representing the pages individually.

\subsubsection{Activity based features $(G_{a})$ }
Earlier research \cite{wilkinson2007assessing} shows that there exists a strong correlation between article quality and the editing activity; an abrupt spike in activity, or very less activity could be potential indicators of quality change. As we discussed in the contribution based features, the editors contribute more before the quality promotion, and as a result, communication and coordination is increased among them. Hence, the number of revisions in the main and the talk page increases, indicating the rise in reverts and undo/redo changes. In contrast, a sudden fall in the number of revisions followed by the stagnancy in further changes may indicate quality demotion. We included various edit activity based features in our model.
\begin{itemize}
    \item \textit{Mean time elapsed (at the granularity of months) between two consecutive revisions of the article main pages.} (\textbf{F7})
    \item \textit{Variance of time elapsed (at the granularity of months) between two consecutive revisions of the article main pages.} (\textbf{F8})
    \item \textit{Mean time elapsed (at the granularity of months) between two consecutive revisions of the article talk pages.} (\textbf{F9})
    \item \textit{Variance of time elapsed (at the granularity of months) between two consecutive revisions of the article talk pages.} (\textbf{F10})
    \item \textit{Number of revisions of the talk pages (at the granularity of months).} (\textbf{F11})
    \item \textit{Number of revisions of the talk page (at the granularity of weeks).} (\textbf{F12})
    \item \textit{Number of revisions of the main page (at the granularity of months).} (\textbf{F13})
    \item \textit{Number of revisions of the main page (at the granularity of weeks).} (\textbf{F14})

\end{itemize}
\subsubsection{Content based features $(G_{p})$}
Previous works have shown that article content plays an important role in assessing quality and hence we have considered the article text, i.e., content of the main namespace as possible indicators of quality change. In particular, we have extracted the following features for every page as reported in \cite{zhang2018history, shen2017hybrid}. 
\begin{itemize}
    \item \textit{Article length in bytes.} (\textbf{F15})
    \item \textit{Number of references.} (\textbf{F16})
    \item \textit{Number of categories mentioned in the text.} (\textbf{F17})
    \item \textit{Number of links to other articles.} (\textbf{F18})
    \item \textit{Number of citation templates.} (\textbf{F19})
    \item \textit{Number of non-citation templates.} (\textbf{F20})
    \item \textit{Number of images/article length.} (\textbf{F21})
    \item \textit{If infobox template exists.} (\textbf{F22})
    \item \textit{Number of level 2 section headings.} (\textbf{F23})
    \item \textit{Number of level 3+ section headings.}
    (\textbf{F24})
    \item \textit{Information noise score.} (\textbf{F25}) (adapted from \citeauthor{zhu2000incorporating}~\cite{zhu2000incorporating})
    \item \textit{Readability scores.} (\textbf{F26 : F34}) : 9 types of readability scores.
\end{itemize}
Readability measures how interpretable an article is to the readers and computes a score based on the use of language in the article. This is used as an important measure to reflect the encyclopedic standard of representing its content. We have included $9$ readability scores in our features as discussed in \cite{shen2017hybrid}.

\noindent
Finally, every revision of an article is represented as a vector of all the above 34 features for the detection of quality change points.

\subsubsection{Correlation among the features}

\begin{figure}[h!]
    \centering
    \includegraphics[height=12cm, width=0.8\columnwidth]{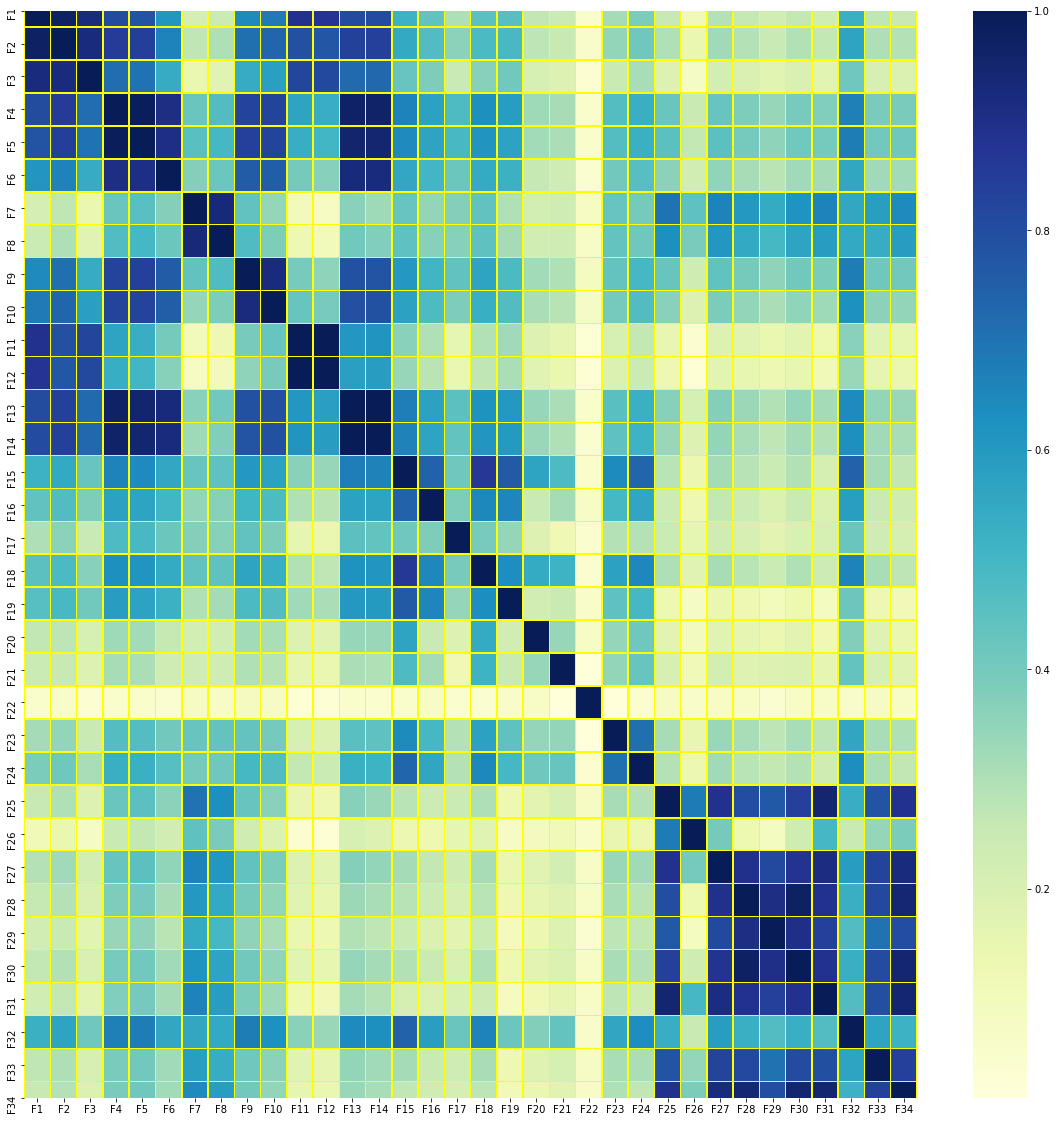}
    \caption{Heatmap showing the correlations between the various categories of features. Blue $\rightarrow$ Sky $\rightarrow$ White indicates 1.0 $\rightarrow$ 0.0 $\rightarrow$ -1.0 correlation values. The heatmap follows the feature order (top to bottom) in which the first 6 features denote the \textit{contribution based features} followed by 8 features of the \textit{activity based features}. The last 20 features in the heatmap define the \textit{content based features}.}
    \label{fig:heatmap}
\end{figure}

\blu{To examine the relationship among the different features ($G_{c}$ , $G_{a}$, $G_{p}$), we compute the Pearson’s correlation coefficients between all possible feature combinations. Unlike the standard form of feature vectors used in prediction models, every feature in our work is represented as the time series of distinct feature values for the time span (i.e., 156 months) we considered. This is true for every individual article.
Formally, let us assume, a feature $F_{i}$ of an article $A_{i}$ is represented as time-series of $m$ timestamps, i.e., spanning over $m$ months. Also, let us denote the $i^\textrm{th}$ timestamp as $T_{i}$ for the feature $F_{i}$. We have 14872 articles (i.e., say $n$) in our dataset that had undergone quality changes at least once in the lifetime. We averaged the feature $F_{i}$ of $n$ articles at timestamp $T_{i}$. We followed the same method for all the timestamps (which is 156 in number) and thus obtained 156 values of $F_{i}$. The above mentioned procedure has been followed for all the features. This results in a matrix of dimension of 34 * 156, in which the number of features and timestamps is 34 and 156 respectively. Figure \ref{fig:heatmap} shows the correlation coefficient matrix for the 34 features.}


\blu{We observe that the features belonging to same category ($G_{c}$ or $G_{a}$ or $G_{p}$) are positively correlated but are negatively correlated with the features of other categories. However almost all the coefficients are in the range -0.08 to 0.08 except for the pairs which belong to the same category of features and are semantically close to each other. For example, the features - number of editors editing the article pages (i.e., F4) and number of new editors editing the article pages (i.e., F5) under \textit{contribution based features $(G_{c})$} or the different readability scores except the feature F26 of \textit{content based features $(G_{p})$} are closely related and hence are showing positive correlation in the heatmap. Also, we found that \textit{content based features $(G_{p})$} except readability scores are less correlated with each other, especially the feature whether infobox template exists (i.e., F22) is negatively correlated within the group.}

\subsubsection{Analysis of temporal changes in the features}
\begin{figure}[h]%
    \centering
        \subfloat[\centering \label{fig:editors_analysis} Number of distinct registered editors editing the article main page]{{\includegraphics[height=2.90cm]{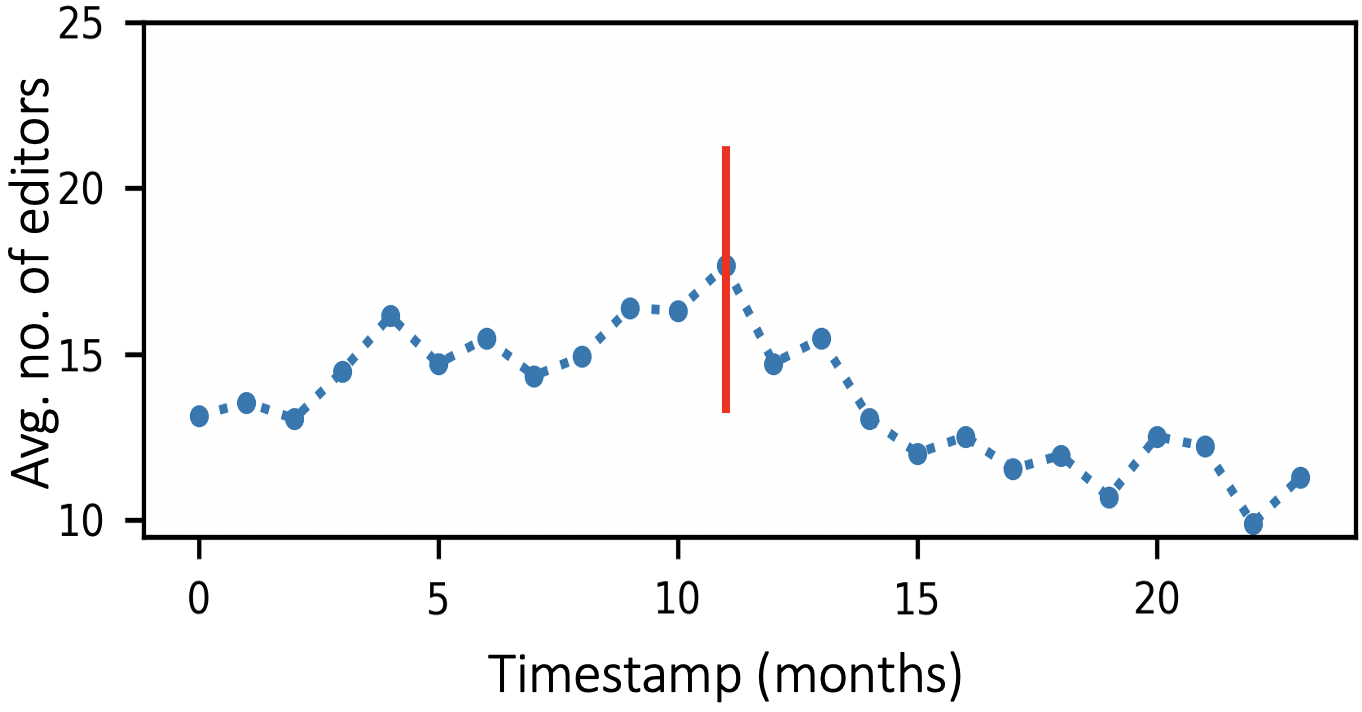}}}%
    \qquad
    \subfloat[\label{fig:revision_analysis}\centering Number of revisions of the main page]{{\includegraphics[height=2.90cm]{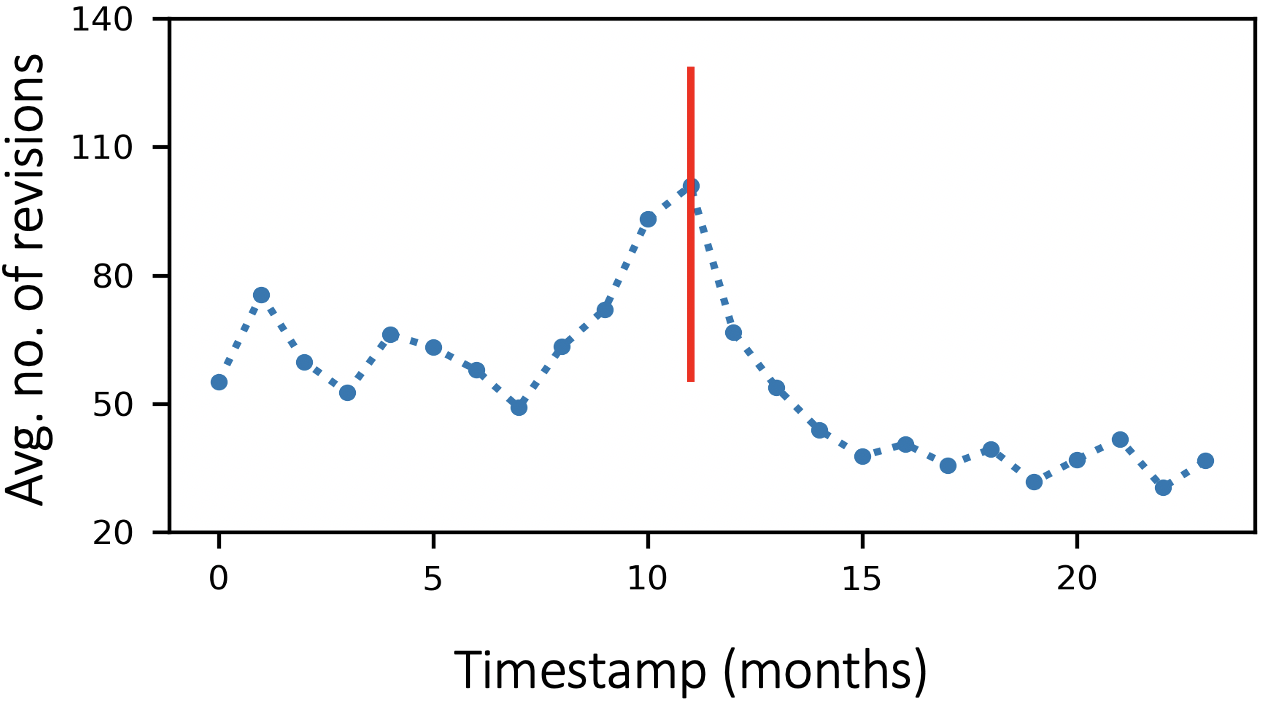}}}%
    \qquad
    \subfloat[\label{fig:length_analysis}\centering Article length in bytes]{{\includegraphics[height=2.90cm]{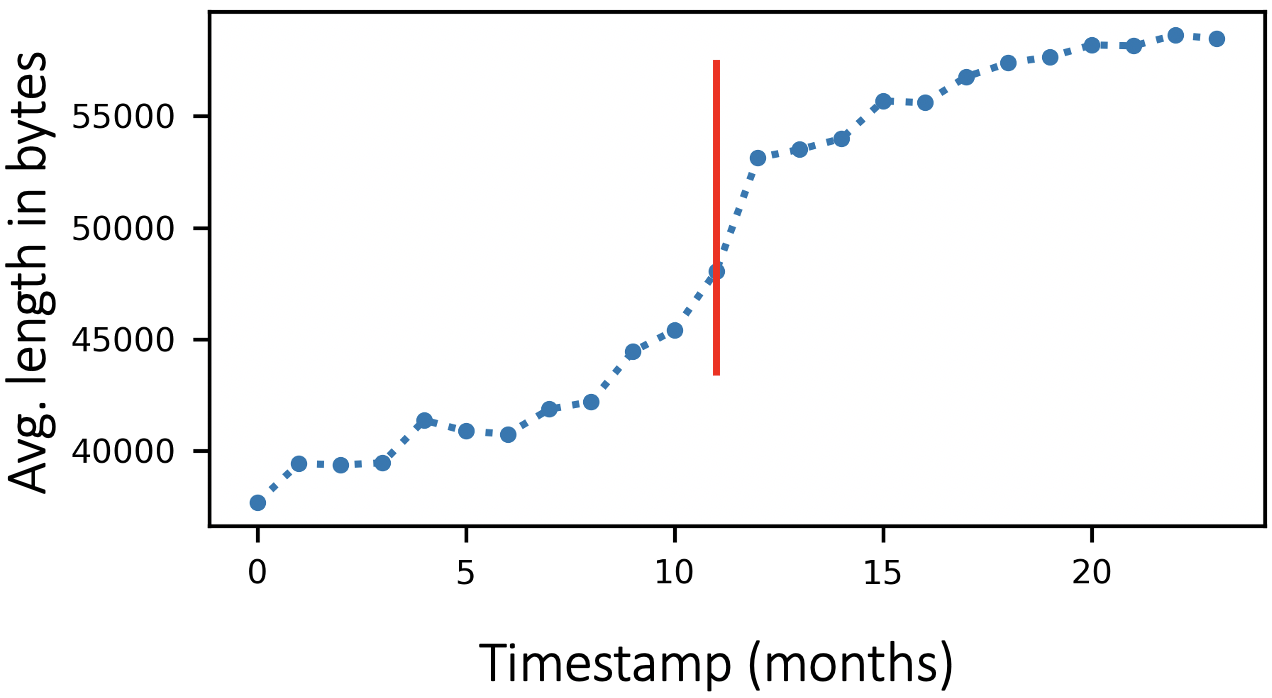}}}

    \caption{The temporal pattern of three features from the three feature categories. The features are computed for 54 pages that have changed from the FA quality to some lower quality and the red line indicates the change point. The data points represent the mean values of the 54 pages in each case.}
    \label{fig:feature_analyses}
\end{figure}

\blu{We have presented a number of quality change patterns and their statistics so far. As a next step we wanted to observe the temporal pattern of changes of the features at the point of the quality change.
In particular we were interested in cases that underwent both \textit{promotion and demotion in quality}, which resulted in a selection of $54$ pages that underwent a change from the FA quality (the highest quality) to some lower quality (AGA, BC, SS) at least once in their life-time.} The time window is set as the $24$ consecutive timestamps (i.e., months) including the change point. The articles that earned the FA quality, undergo several rounds of review and hence abrupt changes in the feature space are expected before and after the quality change (e.g., change from FA to BC). The red line indicates the change point in each of the plots in Figure \ref{fig:feature_analyses}. In every case, the data points represent the mean values of the features of $54$ pages.
\begin{itemize}
    \item As shown in Figure \ref{fig:feature_analyses}(a), the number of distinct registered editors editing the article main page declines suddenly after the quality change point. If we consider $4$ consecutive timestamps on either side of the change point (red line), the change is sharp on both sides. This indicates that a large number of editors were active when the pages reached the FA status but many of them stopped editing (the main page) after the demotion in quality.  
    \item Similarly in Figure \ref{fig:feature_analyses}(b), we plotted the number of revisions of the main page and observed very similar trends as with the number of editors. There is a huge difference in the number of revisions on both sides of the red line, indicating a sharp change in the activity of the community.
    \item In Figure \ref{fig:feature_analyses}(c), the length of the article increases sharply at the point of quality change. The increase in bytes can signal that the editors added more content to the articles after the demotion compared to the earlier phase.
\end{itemize}
Overall, the temporal patterns of the three features show shift in their mean values at the point of quality changes \blu{(see Appendix~\ref{appendix:A} for the temporal trends for all the features)}. The analysis revealed that such patterns could be particularly helpful for the CPD methods to detect the change points (cf section \ref{sec:CPD_methods}). 

\subsection{Change point detection} \label{sec:CPD_methods}

\subsubsection{The time-series settings}
We now briefly describe the problem of \textit{change point detection} (\textit{CPD}) and its resemblance to the quality change detection in our setting. In a {CPD} problem, given a set of observations $\mathcal{Y}$ for time steps $t = 1,2,\dots$, we need to identify the time indexes where there is an abrupt change in the behavior of the time series, specifying a probable signal to the alteration of the data generation process. These indexes are denoted as \textit{change points} and a number of CPD algorithms are used to detect the change points. The algorithms are about to estimate the unknown instants where the characteristics of the series change abruptly. Depending on the context, the change points are used in evaluating the correctness of the algorithm.

In our settings of quality change point detection, we postulate the analogy as follows.
\begin{itemize}
    \item Individual article is assumed as separate time series and the timeline is expressed in months. In other words, we consider the revisions on a monthly basis. 
    \item The sample data points are the latest revision in each month given a typical Wikipedia article.
    \item The change points are the time instants, specifically the months where the quality of the article were changed. We shall be using the terms months and timestamps interchangeably. 
\end{itemize}

Mathematically, let $\mathbf{y}_t\in \mathcal{Y}$ denote the representation of the article at the timestamps $T_t$ $\forall t = 1, 2, \dots, N$. Here, $N$ is the number of months in our timeline. Let us denote the size of the domain of every $\mathbf{y}_t$ by \textit{d}, which is the total number of features described earlier, and therefore $\mathcal{Y}\subset\mathrm{R}^d$. We denote $\mathbf{y}_{i:j}$, $i<j$ as the segment of timestamps $T_i, T_{i+1}, \dots, T_{j}$. The timestamp ordered set of quality change points is denoted by $\mathcal{Q}=\{q_1, q_2, \dots, q_n\}$, where $n$ is the number of quality change points for the given article and used as the ground truth. We use $q_0 = {T_1}$, the timestamp at which the page is created, and $q_{n+1} = T_{N}+1$ as the endpoints of the quality change point set. It is required to note that we add these points only for convenience of evaluation metrics which are defined later in this section, and we do not consider these points as the actual quality change points.

\subsubsection{The CPD algorithms}
We experiment with a few CPD algorithms ranging from one of the earliest proposed models to the more recently proposed ones. Since our feature space is multi-dimensional, we experiment with the multivariate setting only. Furthermore, since our objective is to analyze and infer observations from the features that are used in detecting the quality change points, we have experimented mainly with the offline CPD algorithms for retrospective detection and analysis of the change points. In the offline multivariate change point detection algorithms, the general objective is to optimize the following cost function
\begin{equation}
\centering
    \min\limits_{\mathcal{Q}}\sum_{i=1}^{n+1} \mathcal{L}(\mathbf{y}_{\mathcal{Q}_{i-1}:\mathcal{Q}_i-1}) + \lambda P(n)
\label{equ:cost_function}
\end{equation}
where $\mathcal{L}(.)$ is the loss associated for the segment $\mathbf{y}_{\mathcal{Q}_{i-1}:\mathcal{Q}_i-1}$, $\lambda$ is a hyperparameter, and $P(n)$ is a penalty on the number of change points. The intuition for this function is to minimize the loss in grouping the timestamps with the same quality of the article in a single contiguous segment. Also note that the point at which the quality changes ($\mathcal{Q}_i$) is considered in the new segment. 

For the close compatibility with change point detection and localization in multivariate data with multiple change points, we applied the following change point detection algorithms in our work.
\begin{itemize}
    \item Binary Segmentation [{\em aka} \textbf{BinSeg}] - It is one of the earliest methods for detecting the change points that greedily splits the timestamp series into disjoint segments based on optimising a predefined cost function. For the quality change detection, we use the cost function defined in Equation \ref{equ:cost_function}. The time complexity associated with this method is $O(N log N)$. 
    \item Pruned Exact Linear Time [{\em aka} \textbf{PELT}] - This is an offline method that works through minimising the aforementioned cost function over possible numbers and locations of change points. Through minimization, the approach achieves the optimal number and location of change points that has a computational cost which under mild conditions, is linear time in the number of observations.
    \item Non-parametric Change Point Detection [{\em aka} \textbf{ECP}] - It is a nonparametric approach for detecting the change points. For a set of multivariate observations of arbitrary dimensions, the model performs a nonparametric estimation of both the number of change points and the locations at which they occur. The estimation of the change points is based on hierarchical clustering of the timestamps.
\end{itemize}
\subsection{Evaluation}
For evaluating the performance of the methods, we use the metrics described in \cite{van2020evaluation}. These metrics are compatible with multiple change point (ground truth) setting and also quantify the consistency of the annotations of the timestamps. These metrics can be roughly categorised into clustering and classification metrics. The locations of the ground truth change points are denoted by the ordered set $\mathcal{G} = \{g_1, g_2, \dots, g_k\}{\ }\forall g_i\in\{T_1, T_2, \dots, T_N\}$ and $g_i<g_j$ for $i<j$. The set $\mathcal{G}$ partitions the timestamps into disjoint sets $s_j\in\mathcal{S}$, where $s_j$ is the segment from $T_{j-1}$ to $T_{j}-1\ \forall j\in \{1, 2, \dots, k+1\}$. The clustering based metrics evaluates the CPD algorithms based on the view that change point detection inherently aims to divide the timestamps into distinct regions with a constant quality of the article. 

\noindent\textbf{Change point evaluation as clustering}: Among the different clustering metrics proposed by several algorithms \cite{hubert1985comparing, hausdorff1927mengenlehre, everingham2010pascal, arbelaez2010contour}, we used the \textit{covering metric} because of its ability to show the true performance of methods that report many false positives. For any two sets $s, s'$, the Jaccard index is computed using the following expression
\begin{equation}
    J(s, s') = \frac{|s \cap s'|}{|s \cup s'|}
\end{equation}
The authors in~\cite{arbelaez2010contour} define the covering metric of a partition $\mathcal{S}$ by partition $\mathcal{S}'$ as 
\begin{equation}
    \mathcal{C}(\mathcal{S}, \mathcal{S}') = \frac{1}{N}\sum_{s\in\mathcal{S}}|s|.\min\limits_{s'\in\mathcal{S}'} J(s, s')
\end{equation}
where partition $\mathcal{S}$ is the partition induced by the ground truth set $\mathcal{G}$, and $\mathcal{S}'$ is the partition induced by the set $\mathcal{Q}$ predicted by the model. 

\noindent\textbf{Change point evaluation as classification}: A different set of evaluation metrics for CPD algorithms considers the change point detection as a classification problem between the ``change point" and ``non-change point" classes \cite{killick2012optimal, aminikhanghahi2017survey}. The simple metrics such as accuracy will be highly skewed because the number of quality change points of a page will be very small compared to the total number of revision timestamps of the article. Therefore, we look at the effectiveness of the algorithms in terms of \textit{precision} and \textit{recall}. The set of true positives in the predicted set $\mathcal{Q}$, denoted by $\mathrm{TP}(\mathcal{G}, \mathcal{Q})$, consists of all the timestamps $g\in\mathcal{G}$ for which $\exists q\in\mathcal{Q}$ such that $|g-q|\leq M$, while ensuring that only one $q\in\mathcal{Q}$ is used for one $g\in\mathcal{G}$. The value $M$ is a commonly defined margin of error around the true change point location to allow for minor discrepancies, which is an usual practice in evaluating the change point detection algorithms~\cite{killick2012optimal, truong2020selective, van2020evaluation}. However, the additional condition imposed avoids double counting, so that among the multiple detection within the margin around a true change point only one is recorded as a true positive \cite{killick2012optimal}. The precision and recall are then defined as follows. 
\begin{align}
    P = \frac{|\mathrm{TP}(\mathcal{G}, \mathcal{Q})|}{|\mathcal{Q}|}\\
    R = \frac{|\mathrm{TP}(\mathcal{G}, \mathcal{Q})|}{|\mathcal{G}|}
\end{align}
According to this definition the false positives are the ones that do not have any corresponding ground truth points.

\begin{figure}
    \centering
    \includegraphics[width=\textwidth,height=13cm]{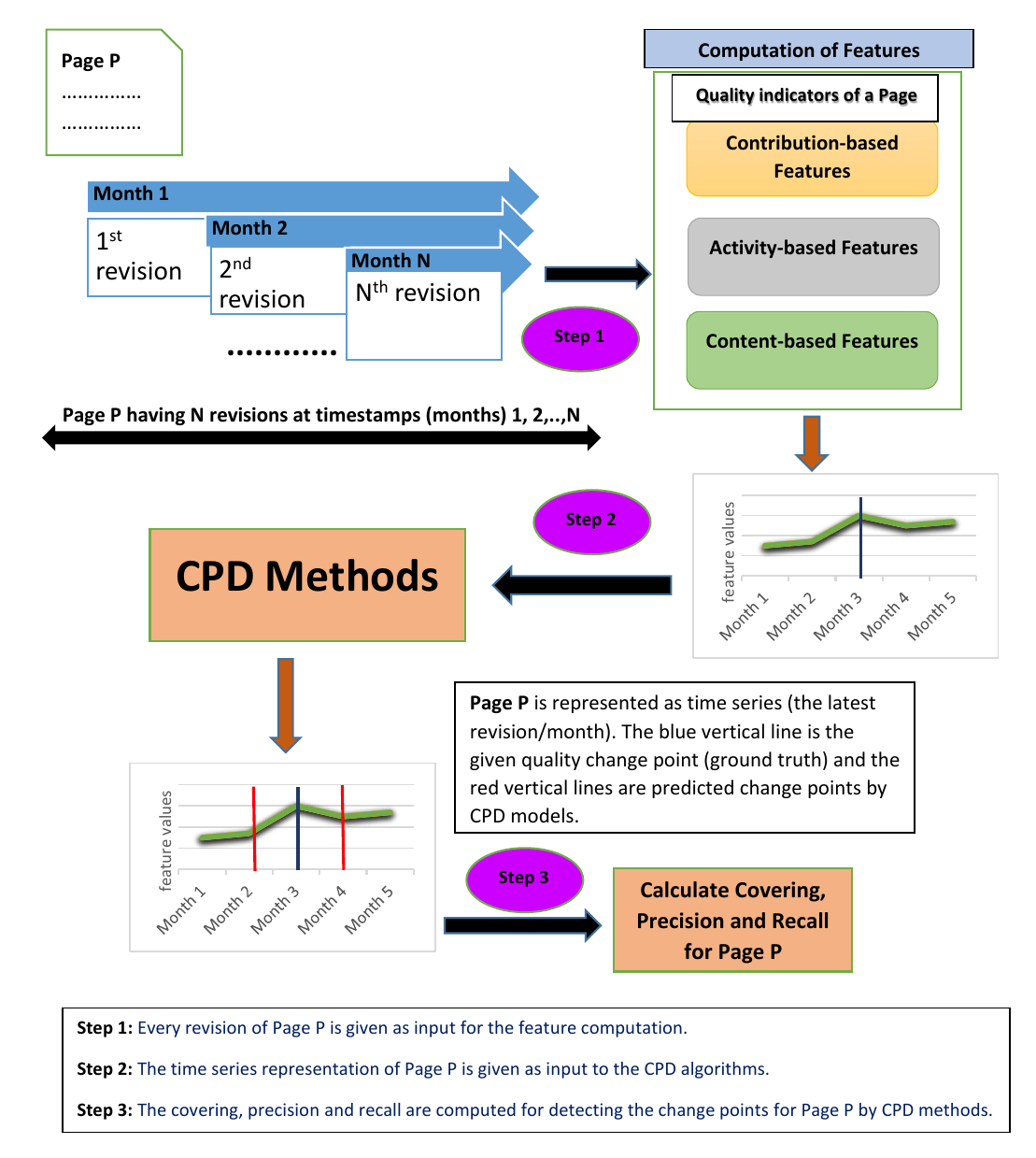} 
   \caption{The pipeline for unsupervised change point detection for an individual article.}
   \label{fig:flow1}
\end{figure}

\section{Experiment and Results} \label{sec:res}
\blu{
We run the three algorithms -- \textbf{BinSeG}, \textbf{PELT} and \textbf{ECP} on a sample of our dataset (discussed in the next section) for all possible combinations of the mentioned features. In addition, we also present an ensemble of the three algorithms, which we term as the \textbf{HYBRID} algorithm. This simply takes the best of the three methods and outputs the same. Further, inspired by the feature correlation analysis in the previous section, we present a number of ablation studies to understand which features are the strongest determinants of quality change points.}   

\if{0}\subsection{Feature correlation}
\textcolor{blue}{
To examine the relationship among the different features ($G_{c}$ , $G_{a}$, $G_{p}$), we compute the Pearson’s correlation coefficients between all possible combination-pair of features. Unlike the standard form of feature vectors used in prediction models, every feature in our work is represented as the time series of distinct feature values for the time span (i.e., 156 months) we considered. This is true for every individual article. Formally, let us assume, a feature $F_{i}$ of an article $A_{i}$ is represented as time-series of $m$ data samples, spanned over $m$ months. We represented the feature $F_{i}$ as the average of $m$ values for the article $A_{i}$. We followed the same method for all the articles ( which is 14872 in number) and thus the feature $F_{i}$ is presented as a list of 14872 values. The above mentioned procedure has been followed for all the features. This results a matrix of dimension of 34 * 14872, in which the number of features and articles is 34 and 14872 respectively. Figure \ref{fig:heatmap} shows the correlation coefficient matrix for the 34 features. We observe that the features belonging to same category ($G_{c}$ or $G_{a}$ or $G_{p}$) are positively correlated but are negatively correlated with the features of other categories. However almost all the coefficients are in the range -0.08 to 0.08 except for the pairs which belong to the same category of features and are semantically close to each other. For example, the features- number of editors editing the article pages (i.e., F4) and number of new editors editing the article pages (i.e., F5) under \textit{Contribution based features $(G_{c})$} or the different readability scores except the feature F27 of \textit{Content based features $(G_{p})$} are closely related and hence are showing positive correlation in the heatmap. Also, we found that \textit{Content based features $(G_{p})$} except readability scores are less correlated with each other, especially the feature whether infobox template exists (i.e., F22) is negatively correlated within the group.}\fi

\subsection{Experimental setup}
For preparing the sample set to run the algorithms, we consider only those articles that have more than one ground truth quality change points. This leaves us back 14872 out of 30826 articles. 
This is since it is essential for the CPD methods to have at least one ground truth point to distinguish the change in data generation process of the time series. 
Every article was represented as a distinct time series of revisions at different timestamps and each revision as a $d$ dimensional feature vector. 
We consider only the latest revision in a month to filter out the noise in the dataset that occurs due to the quality switch war which has been explained before (cf section \ref{sec:Cyclic}). We have experimented with various time intervals ranging from ten days to six months and have observed that by using one month we retain 95\% of the quality change points, which also resulted in the best results. Since all features we use are cumulative in nature, this choice does not affect the feature computation process.

\blu{To investigate the performance of CPD algorithms, we divided the dataset comprising of 14872 articles into two sample sets -- train and test sets. We divided the dataset comprising  14872 articles into two sample sets - training and test sets. The train-test split for our experiment is set to 80:20. We tried to maintain uniform distribution of articles of different quality classes across the two sets so that the results do not get dominated by the performance of a single class containing a specific set of articles. For achieving the best results, we tuned the hyperparameters of the CPD algorithms on the training set and later tested them on the test set. A schematic diagram of our experimental setup for the change point detection is illustrated in Figure \ref{fig:flow1}.}


\subsection{Hyperparameter settings}
\blu{The penalty value, i.e, \textit{pen\_val} of the PELT and \textit{n\_bkps} of the BinSeG algorithm, which determines the significance of the change points identified, is set to 1, which performed best in the range of values from 1 to 8. The mean number of change points (i.e., ground truth) observed for the articles in the training set is 2.63. For both PELT and BinSeG methods, we use rbf to model the cost function. For the ECP algorithm, we varied the \textit{min\_size} parameter, which defines the minimum gap between two successive change points from the set of values 2, 5, 10, 15, 20. The best results are observed at 5. The optimal value of the hyperparameters are observed for the articles in the training set. For evaluation in terms of precision and recall, the margin of error $M$ is set to 5 ($\pm$ 5 on either side of the predicted point). We ran the three CPD algorithms with their optimal hyperparameters on every test page individually and finally aggregated metric values over all the test samples. For the HYBRID method, we took the maximum one reported by the three algorithms in their best hyperparameter settings. The mean values for each of the metrics, i.e., covering, precision, recall for all the combinations of the features are reported in the Table \ref{tab:result_alldata}}.


\subsection{Key results} 
We achieved the best covering of \textbf{0.76} from BinSeG. The best result for ECP and PELT are 0.6. As expected, the content based features alone are sufficient to produce the best results (see~\cite{Maity2015AnalysisAP} for similar observations). The highest achievable precision and recall are \textbf{0.6} (BinSeG) and \textbf{0.61} (PELT) respectively. Once again these best results are achieved using just the content based features. Overall among the three algorithms, PELT achieves the best compromise considering all the three results (covering $=0.6$, precision $=0.37$ and recall $=0.61$). The HYBRID method, as it should be, achieves the best covering ($0.79$), precision ($0.68$) and recall ($0.69$) for the content based features.

\begin{table*}[h]
    \centering
    \scalebox{0.68}{
    \begin{tabular}{|c|r|r|r|r|r|r|r|r|r|r|r|r|}
    \hline
    \multirow{3}{*}{Features} &
    \multicolumn{3}{c|}{BinSeG [n\_bkps = 1]}  & \multicolumn{3}{c|}{ECP [min\_size = 5]} &
    \multicolumn{3}{c|}{PELT [pen\_val = 1]} &
    \multicolumn{3}{c|}{HYBRID} \\
    \cline{2-13}
    & Covering & Precision & Recall & Covering & Precision & Recall & Covering & Precision & Recall & Covering & Precision & Recall \\ 
    \hline
    $G_{c}$ & 0.68 & 0.36 & 0.29 & 0.39 & 0.23 & 0.27 & 0.52 & 0.28 & 0.47 & 0.73 & 0.51 & 0.60\\
    \hline
    $G_{a}$ & 0.68 & 0.37 & 0.30 & 0.41 & 0.27 & 0.29 & 0.52 & 0.29 & 0.47 & 0.74 & 0.53 & 0.59\\
    \hline
    $G_{p}$ & \cellcolor{green!20}0.76 & \cellcolor{green!20}0.60 & 0.46 & 0.60 & 0.45 & 0.45 & 0.60 & 0.37 & \cellcolor{green!20}0.61 & \cellcolor{blue!20}0.79 & \cellcolor{blue!20}0.68 & \cellcolor{blue!20}0.69\\
    \hline
    $G_{c} \oplus G_{a}$ & 0.68 & 0.38 & 0.30 & 0.42 & 0.27 & 0.31 & 0.53 & 0.29 & 0.45 & 0.74 & 0.53 & 0.59\\
    \hline
    $G_{a} \oplus G_{p}$ & 0.75 & 0.56 & 0.43 & 0.59 & 0.42 & 0.44 & 0.60 & 0.37 & 0.58 & 0.78 & 0.66 & 0.68\\
    \hline
    $G_{p} \oplus G_{c}$ & 0.74 & 0.53 & 0.41 & 0.56 & 0.39 & 0.43 & 0.60 & 0.37 & 0.57 & 0.78 & 0.64 & 0.60\\
    \hline
    $G_{c} \oplus G_{a} \oplus G_{p}$ & 0.74 & 0.53 & 0.41 & 0.56 & 0.39 & 0.43 & 0.60 & 0.37 & 0.57 & 0.78 & 0.64 & 0.67\\
    \hline
    \end{tabular}}
    \vspace{2mm}
    \caption{CPD outcome: A comparison of the BinSeG, ECP and PELT algorithms on test set. Best results are highlighted in \textcolor{green}{green}. Results highlighted in \textcolor{blue}{blue} are the best among those achieved by the \textbf{HYBRID} method.}
    \label{tab:result_alldata}
\end{table*}

\subsection{Additional experiments}
Further, we analyzed the obtained results based on two special criteria.
\begin{itemize}
    \item \textbf{Criteria 1:} We considered only those articles that have at least three change points (i.e., assuming that larger number of change points on the time series should help in having a better inference). This makes $1982$ articles in this category.
    \item \textbf{Criteria 2:} Here we considered only those articles whose latest class is FA ($3190$ articles). This was precisely to understand how well the model performs for the highest quality and possibly the most important class. 
\end{itemize}
We run the ECP algorithm on \blu{the test samples belonging to the mentioned criteria 1 and criteria 2} (other algorithms produce similar results and hence, not shown) and obtain the values of the three evaluation metrics. \blu{The train-test split (80:20) and hyperparameter values remain same as described earlier (see section 6.2).} The results are noted in the Table~\ref{tab:result_criteria}. As is expected, the results are considerably better than the entire dataset for all the feature combinations.

\begin{table}[h!]
    \centering
    \scalebox{0.75}{
    \begin{tabular}{|c|r|r|r|r|r|r|}
    \hline
    \multirow{3}{*}{features} & 
    \multicolumn{3}{c|}{Criteria 1} & 
    \multicolumn{3}{c|}{Criteria 2} \\
    \cline{2-7}
    & Covering & Precision & Recall & Covering & Precision & Recall \\
    \hline
    $G_{c}$ & 0.53 & 0.39 & 0.33 & 0.52 & 0.37 & 0.37\\
    \hline
    $G_{a}$ & 0.58 & 0.48 & 0.40 & 0.56 & 0.44 & 0.44\\
    \hline
    $G_{p}$ & \cellcolor{green!20}0.72 & \cellcolor{green!20}0.67 & 0.47 & \cellcolor{green!20}0.71 & \cellcolor{green!20}0.65 & \cellcolor{green!20}0.56\\
    \hline
    $G_{c} \oplus G_{a}$ & 0.59 & 0.48 & 0.41 & 0.57 & 0.43 & 0.44\\
    \hline
    $G_{a} \oplus G_{p}$ & 0.71 & 0.60 & \cellcolor{green!20}0.50 & 0.69 & 0.58 & \cellcolor{green!20}0.55\\
    \hline
    $G_{p} \oplus G_{c}$ & 0.69 & 0.58 & 0.46 & 0.67 & 0.55 & 0.53\\
    \hline
    $G_{c} \oplus G_{a} \oplus G_{p}$ & 0.67 & 0.56 & 0.48 & 0.65 & 0.52 & 0.53\\
    \hline
    \end{tabular}
    }
    \caption{CPD outcome of ECP algorithm (other algorithms show similar trends and hence not shown) for the two different special criteria. Best results for each criteria are highlighted in \textcolor{green}{green}.}
    \label{tab:result_criteria}
\end{table}

\subsection{Ablation study}
\blu{To understand the contribution of specific features to the outcome of the CPD algorithms, we performed several ablation studies of feature combinations of three specified categories - $G_{c}, G_{a}$ and $G_{p}$. The results show that content based features $G_{p}$ achieves the best performance. We performed an ablation study in which one feature was masked and the CPD algorithm was run for the remaining features. In every step the drop in the values of three metrics are noted. The drop is not significant for masking any single feature. We have reported the top 3 changes in values for masking three features in the Table \ref{tab:feature_analysis_mask}. While we show the results for PELT (with \textit{pen\_val} set to 1), the other algorithms show similar trends.}

\begin{table*}[h!]
    \centering
    \scalebox{0.8}{
    \begin{tabular}{|c|c|c|c|}
    \hline
    \multirow{3}{*}{Feature} &
    \multicolumn{3}{c|}{PELT [pen\_val = 1]} \\
    \cline{2-4}
    & Covering & Precision & Recall  \\ 
    \hline
     $G_{c}- F1$ & 0.51 & 0.27 & 0.45\\
     \hline
     $G_{a}- F14$ & 0.53 & 0.28 & 0.45 \\
     \hline
     $G_{p}- F32$ & 0.58 & 0.34 & 0.55\\
     \hline
    \end{tabular}
    }
    \caption{CPD (PELT) Outcome: Results for masking three features individually. Three features that are selected from each category of $G_{c}, G_{a}, G_{p}$ are (i) Number of registered editors editing talk pages, (ii) Number of revisions of article page per week and (iii) difficult words (readability score) respectively. }
    \label{tab:feature_analysis_mask}
\end{table*}

\blu{Next, we tried combination of different features irrespective of the feature category and ran the CPD algorithm. The choice of the features were motivated by the correlation coefficients as shown in the heatmap (see Figure \ref{fig:heatmap}). We have used the abbreviations for the combination of features as mentioned below. The notations of the features are same as in the heatmap (see Figure \ref{fig:heatmap}).
\begin{itemize}
    \item \textbf{G1:} All the readability features, i.e., {F26 : F34}
    \item \textbf{G2:} All the content based features, i.e., F15 to F25 except the readability features.
    \item \textbf{G3:} All the features of G2 and the readability feature F32.
    \item \textbf{G4:} A subset of content based features, F15 to F21 and the features, F23 and F24.
    \item \textbf{G5:} All the features of G4 and the readability feature F32.
    \item \textbf{G6:} A subset of activity features, F9 to F14.
    \item \textbf{G7:} The features mentioned in G6 and the readability feature F32.
    \item \textbf{G8:} The features of G7 and all the contribution based features.
\end{itemize}}

\blu{We followed the similar train-test split in tuning the hyperparameters and the result is tabulated in the Table \ref{tab:result_ablation}. Once again we used PELT with the hyperparameter \textit{pen\_val} set to 1.}

\blu{While in Table~\ref{tab:result_alldata} we observe that the best performance comes from the content based features when we dive deeper we observe that a selected combination G7 of activity and content based features chosen as per the heatmap in Table~\ref{tab:feature_analysis_mask} gives superior performance. In specific the features that play instrumental role are 
\begin{itemize}
    \item the mean and the variance of time elapsed between two consecutive revisions of the article talk pages (at the granularity of months),
    \item the number of revisions of the article talk pages (at the granularity of both months and weeks), 
    \item the number of revisions of article main pages (both at the granularity of months and weeks), and.
    \item presence of ``difficult words'' (a readability feature that is built based on the (non) understandability of an undergraduate student in the USA). Examples of some highly frequent ``difficult words'' are `xenon', `pipeline',  `anole', `touchdown', `epilepsy', `carfilzomib'. 
\end{itemize}} 
\blu{Further we see that G3 and G5 feature groups (i.e., the content features article length (in bytes), number of references, number of categories mentioned in the text, number of links to other articles, number of citation templates, number of images/article length, whether infobox template exists, number of level 2 section headings, number of level 3+ section headings and the anti-correlated readability feature ``difficult words'' (among all the readability features) together brings the highest gain in covering. To summarize a set of judiciously selected feature combinations (based on the feature correlations), e.g., mean and variance of time elapsed between two consecutive revisions and the number of revisions of the article talk pages which typically correspond to \textit{the organisational attributes} of the peer-production system and certain \textit{readability attributes} like the presence of ``difficult words'' become crucial when we perform an in-depth analysis which do not manifest in the aggregate level results (i.e., in Table~\ref{tab:result_alldata}).}

\if{0}\blu{To understand the contribution of specific features to the outcome of the CPD algorithms, we performed several ablation studies of feature combinations of three specified categories- $G_{c}, G_{a}$ and $G_{p}$ on the entire dataset (as shown in Table \ref{tab:result_alldata}). Further we conducted the study on the sample subsets of articles belong to two special criteria (see Table \ref{tab:result_criteria}). In both the cases, the result shows that content based features $G_{p}$ reported the best outcome. 
In this section, we tried combination of different features irrespective of the feature category. The experiments were conducted on two sets of articles- (i) all the articles, (ii) articles of individual quality classes, i.e., FA, AGA and BC, SS. The result are tabulated in the Table \ref{tab:result_ablation} and Table \ref{tab:result_ablation_special} respectively.
The choice of the features depend on the correlation coefficient as shown in the heatmap (see Figure \ref{fig:heatmap}). We implemented the CPD algorithm PELT with the hyperparameter \textit{pen\_val} set to 1 as the optimal choice of hyperparmeter setting. 
We have used the abbreviations for the combination of features as mentioned below. The order (i.e., index) of the features followed the same as shown in the heatmap (see Figure \ref{fig:heatmap}) in bottom-up fashion.
\begin{itemize}
    \item \textbf{G1:} All the readability features, i.e., {F26 : F34}
    \item \textbf{G2:} All the content based features, i.e., F15 to F24 except the readability features.
    \item \textbf{G3:} All the features of G2 and the readability feature F32.
    \item \textbf{G4:} A subset of content based features, F15 to F21 and the features, F23 and F24.
    \item \textbf{G5:} All the features of G4 and the readability feature F32.
    \item \textbf{G6:} A subset of activity features, F9 to F14.
    \item \textbf{G7:} The features mentioned in G6 and the the readability feature F32.
    \item \textbf{G8:} The features of G7 and all the contribution based features.
\end{itemize}
}\fi

\begin{table*}[h]
    \centering
    \scalebox{0.8}{
    \begin{tabular}{|c|r|r|r|}
    \hline
    \multirow{3}{*}{Features} &
    \multicolumn{3}{c|}{PELT [pen\_val = 1]} \\
    \cline{2-4}
    & Covering & Precision & Recall  \\ 
    \hline
     G1 & 0.53 & 0.32 & 0.74\\
     \hline
     G2 & 0.52 & 0.28 & 0.37 \\
     \hline
     G3 & \cellcolor{green!20}0.61 & 0.39 & 0.57\\
     \hline
     G4 & 0.52 & 0.27 & 0.37 \\
     \hline
     G5 & \cellcolor{green!20}0.62 & 0.39 & 0.57\\
     \hline
     G6 & 0.53 & 0.30 & 0.51\\
     \hline
     G7 & \cellcolor{green!20}0.62 & \cellcolor{green!20}0.41 & \cellcolor{green!20}0.69 \\
     \hline
     G8 & 0.60 & 0.39 & 0.63\\
     \hline
    \end{tabular}
    }
    \caption{CPD (PELT) outcome: Results for different combination of features on test data. Best results are highlighted in \textcolor{green}{green}.}
    \label{tab:result_ablation}
\end{table*}

\blu{In order to further understand the classwise importance of features, we report the performance metrics for each class in Table \ref{tab:result_ablation_special}. This based on the feature groups introduced earlier (i.e., G1--G7) and the PELT model.  An universal trend is that the best covering is obtained for the feature group G5. Overall, for the majority of the performance metrics, G7 is the winner thus pointing to the universality and robustness of our results.}

\begin{table*}[h]
    \centering
    \scalebox{0.68}{
    \begin{tabular}{|c|r|r|r|r|r|r|r|r|r|}
    \hline
    \multirow{3}{*}{Features} &
    \multicolumn{3}{c|}{FA class}  & \multicolumn{3}{c|}{AGA class } &
    \multicolumn{3}{c|}{BC and SS class} \\
    \cline{2-10}
    & Covering & Precision & Recall & Covering & Precision & Recall & Covering & Precision & Recall \\ 
    \hline
    G1 & 0.56 & 0.38 & 0.77 & 0.56 & 0.37 & 0.80 & 0.49 & 0.25 & 0.70 \\
    \hline
    G2 & 0.61 & 0.39 & 0.48 & 0.54 & 0.32 & 0.41 & 0.47 & 0.17 & 0.28 \\
    \hline
    G3 & 0.61 & 0.39 & 0.48 & 0.54 & 0.32 & 0.41 & 0.55 & 0.25 & 0.44 \\
    \hline
    G4 & 0.61 & 0.39 & 0.48 & 0.54 & 0.32 & 0.41 & 0.47 & 0.17 & 0.28\\
    \hline
    G5 & \cellcolor{green!20}0.70 & 0.51 & 0.67 & \cellcolor{green!20}0.65 & 0.47 & 0.65 & \cellcolor{green!20}0.55 & 0.25 & 0.44 \\
    \hline
    G6 & 0.59 & 0.39 & 0.60 & 0.55 & 0.34 & 0.54 & 0.46 & 0.18 & 0.41 \\
    \hline
    G7 & 0.69 & \cellcolor{green!20}0.52 & \cellcolor{green!20}0.76 & \cellcolor{green!20}0.65 & \cellcolor{green!20}0.48 & \cellcolor{green!20}0.75 & \cellcolor{green!20}0.55 & \cellcolor{green!20}0.27 & \cellcolor{green!20}0.56 \\
    \hline
    G8 & 0.66 & 0.49 & 0.73 & 0.63 & 0.45 & 0.69 & 0.53 & 0.25 & 0.51 \\
    \hline
    \end{tabular}}
    \vspace{2mm}
    \caption{CPD outcome: A comparison of different combination of features on different quality articles. Best results are highlighted in \textcolor{green}{green}.}
    \label{tab:result_ablation_special}
\end{table*}

\if{0}\textcolor{blue}{
In case of PELT (as shown in Table \ref{tab:result_alldata}), we observed the best result (covering = 0.60, precision = 0.37 and recall = 0.61) was achieved for the content based feature. Although the combination of content and activity based features results the similar covering and precision, the recall is reported lower. We tried to investigate these feature combination at more granular level on the basis of correlation among features. In Table \ref{tab:result_ablation}, we reported covering (0.62), precision (0.41) and recall (0.69) for the feature combination G7 which attains higher score compared to the content based features individually. Similarly, for the fine-grained feature combination of content based features, the highest achievable covering is reported as 0.62 for the feature combination G3 and G5.
We have shown the outcomes of the feature combinations, G1 to G7 in case of three quality classes separately in Table \ref{tab:result_ablation_special}. Here we achieved best covering 0.70, 0.65 and 0.55 for the articles of FA, AGA and BC, SS class for the feature combination G5. 
Although G5 comprises of a subset of content based features only, the combination of activity and content based features in G7 reports the best precision and recall in FA and the best results in all the three metrics for AGA and BC, SS class. This result has opened up new choices in representing articles in terms of organizational as well as textual features.}\fi

\section{Comparison of CPD methods with ORES}\label{sec:ores}
Likewise the popular social media platforms, Facebook, YouTube, Twitter and other corporate and government organizations, Wikipedia has a significant number of AI-based resources that help the  community to take decisions at large scale. Among the pool of bots, human-in-the-loop assisted tools, expert systems, \textit{ORES}\footnote{\url{https://www.mediawiki.org/wiki/ORES}} is a web-service or API, designed for Wikimedia projects to provide an automated solution to critical wiki-works, for example, predicting edit quality, article quality etc. ORES is trained with a large number of machine learning classifiers, operating in real time on the Wikimedia foundation's backend servers and can output a quality score for the given edit or page as a query to the service. In this section we compare the performance of ORES with our CPD algorithms. 

\subsection{The experiment}
For a typical article, we provided every revision to ORES as query and ORES predicted one of the six quality classes. If the predicted class is different from the previous revision, we marked the revision as a predicted quality change point. In such a way, we got the predicted outputs for every revision of a page which is analogous to the outcomes of CPD algorithms. Now, to compute the precision, recall and covering, we assumed the margin of error $M$ same as that of the CPD predictions. Similar to our method, if the predicted change point from ORES is within the margin of error of the ground truth (the change points marked by the editors in the talk pages), the point is marked as a true positive. This allowed us to compute all the metrics -- covering, precision and recall in this setting. As we shall see, the quality indicators (features) and the classification strategy (CPD algorithms vs machine learning classifiers) used in the two settings produced the difference of correctness in prediction task.



\subsection{Result}

We performed the comparison experiments for the two subsets of pages and the results are tabulated in the Table \ref{tab:table_9}. In both cases, we compared the HYBRID method of CPD with ORES. \blu{The optimal hyperparameter values remain same as decided earlier.}
\begin{itemize}
    \item Set 1: Set of pages that have at least one change point in the ground truth reality.
    \item Set 2: The set articles whose latest class is assigned to FA quality. 
\end{itemize}
We observe that our HYBRID CPD algorithm by far outperforms the results obtained from ORES that works on an ensemble of machine learning classifiers. 
\begin{table}[h]
    \centering
    \scalebox{0.8}{
    \begin{tabular}{|c|r|r|r|r|r|r|}
    \hline
    \multirow{3}{*}{\#Articles} &
    \multicolumn{3}{c|}{CPD HYBRID method} &
    \multicolumn{3}{c|}{ORES} \\
    \cline{2-7}
    & Covering & Precision & Recall & Covering & Precision & Recall\\
    \hline
      Set 1 & \cellcolor{green!20}0.79  & \cellcolor{green!20}0.68 & \cellcolor{green!20}0.69 & 0.56 & 0.31 & 0.60\\
    \hline
    Set 2  & \cellcolor{green!20}0.82 & \cellcolor{green!20}0.80 & \cellcolor{green!20}0.73 & 0.63 & 0.40 & 0.71\\
    \hline
    \end{tabular}
    }
    \caption{Performance comparison of HYBRID method with ORES on the test split.}
    \label{tab:table_9}
\end{table}

\section{Discussion} \label{sec:discussion}
\subsection{Further insights from study}
\noindent\textbf{Article quality life-cycle}:
This extraordinary growth of peer-production platforms has sparked the interest of a large body of social computing researchers, primarily to understand the temporal evolution of the co-production system~\cite{arazy2016turbulent,kane2014emergent,keegan2016analyzing,jackson2018folksonomies,arazy2020emergent}. The main goal of this exploration is to identify how quality peer production emerges in the absence of explicit coordination mechanisms. Similarly, in case of Wikipedia, it has been observed that domain specific norms, policies and social roles~\cite{panciera2009wikipedians} emerge organically to facilitate quality content creation. \citeauthor{warncke2015success} ~\citep{warncke2015success} analyzed several quality improvement projects in English Wikipedia and showed that the article quality and, in turn, the success of such projects is associated with certain collaboration archetype and organization - people, purpose and policies. Similarly, we studied all the $30826$ articles individually to investigate the temporal trajectory of quality assigned to the articles which we termed as \textit{article quality life-cycle}. Through our rigorous micro-level analysis of these articles along with the complete article edit log (i.e, text page and talk page both) starting from the creation of the page to its form as of June $2019$, we empirically established that quality change of Wikipedia articles is \textit{more of an exception than a norm}.

Our longitudinal analysis spanning over $14$ years from $2006-2019$ indicates that sequential quality change of article which is expected in participatory knowledge creation paradigm~\cite{Wiki} is observed in only a relatively low fraction of articles. We found that  $51.73\%$ of the articles do not have subsequent quality change after the first shift and this trend is becoming more and more prevalent over the years. Our analysis also demonstrates that lower quality articles have $64.61\%$ higher stagnancy compared to the higher quality articles (cf Section~\ref{sec:temporal_analysis}). \if{0}We demarcate quality changes into $4$ mutually exclusive buckets unlike \cite{Wiki} to circumvent data sparsity problems and, thereby, having statistically meaningful results. The mapping we use is \textbf{FA}, (A+GA)$\rightarrow$\textbf{AGA}, (B+C)$\rightarrow$\textbf{BC} and (Stub+Start)$\rightarrow$\textbf{SS}.\fi We show that out of all articles where quality change is observed, $42.97\%$, show only promotions to higher quality class. Among these changes lower scale promotions such as {\textbf{SS $\Rightarrow$ BC}} involve higher turnaround times, specifically $4$ times higher, compared to higher scale promotions such as {\textbf{BC $\Rightarrow$ AGA}}. This refers to the lack of decisiveness or agency in making ground level administrative decisions for quality assessment. We studied quality changes where only demotions are observed as well as pages where both promotions and demotions take place. We find demotions have $3$ times lower turnaround time compared to promotions. This indicates that any departure from encyclopedic guidelines are detected sooner compared to the time taken to form consensus on conformity to guidelines. Another novel quality shift category which we discover in this work is \textit{cyclic shifts} where article quality oscillates cyclically with consecutive chains of promotions and demotions sometimes revisiting the original state. These quality change actions have low time lags which indicate a parallel to the well known anomalous phenomenon found in Wikipedia articles called {\em edit wars}~\cite{sumi2011edit}.

\noindent\textbf{Implication of dynamic temporal patterns}: Being the largest knowledge repository  Wikipedia community produces at a rate of over 1.9 edits per second. But the quality of this massive production is not a stable attribute and follows a dynamic trajectory impacting the content quality. Because of its non-hierarchical organization Wikipedia trusts on the manual assessments of articles by the editors. Our quantitative temporal analysis has revealed the potential time delay between the content generation and content assessment process which is not favourable for the peer-review platform. Also, besides the collaborative effort by the large number of editors, their coordination and communication play a significant role in the quality maintenance task. Our work has identified instances of conflict and consensus among the contributors/reviewers leading to the diverse dynamic trends in the quality life-cycle of articles. This we believe is an important finding since Wikipedia is not a simple and open environment of collective intelligence, and hence the varying nature of participatory organization, peer-to-peer communications are the influential factors in maintaining the encyclopedia quality broadly.

\noindent\textbf{Quality change points}:
We leveraged the diverse collaborative characteristics of Wikipedia for the automatic prediction of quality changes. We have considered the interplay between the contributors' participation and their agreement statistics in the quality change point detection framework (i.e, $G_{c}$ and $G_{a}$ respectively as quality indicators) along with the article content based features ($G_{p}$). \blu{In general it is true that the content-based features are one of the most predictive ones. This may be attributed to the fact that quality ratings of a Wikipedia article attempt to give an aggregate assessment of different aspects regarding the article such as the coverage of the central topic, organization of the content, the writing style, incorporation of additional resources supporting the verifiability of facts etc. All these aspects fall under the category of content based features and is also observed in prior works. However, as we deep dive into the feature space, we observe that certain combinations of content and activity features show competitive (and in fact better) performance. In particular, presence of ``difficult words'' (a readability feature) in the article seem to be a strong indicator of quality change. Similarly, the activities on the talk pages, e.g., number of revisions, time elapsed between two revisions (i.e., the organisational attributes of Wikipedia) seem to be particularly important. This trend persists across all the quality classes and is universal (see Table~\ref{tab:result_ablation}). Such in-depth analysis and the effects of pairing of content and organisational features was hitherto missing.}

\begin{figure}[h]
    \centering
    \includegraphics[width=1.0\columnwidth]{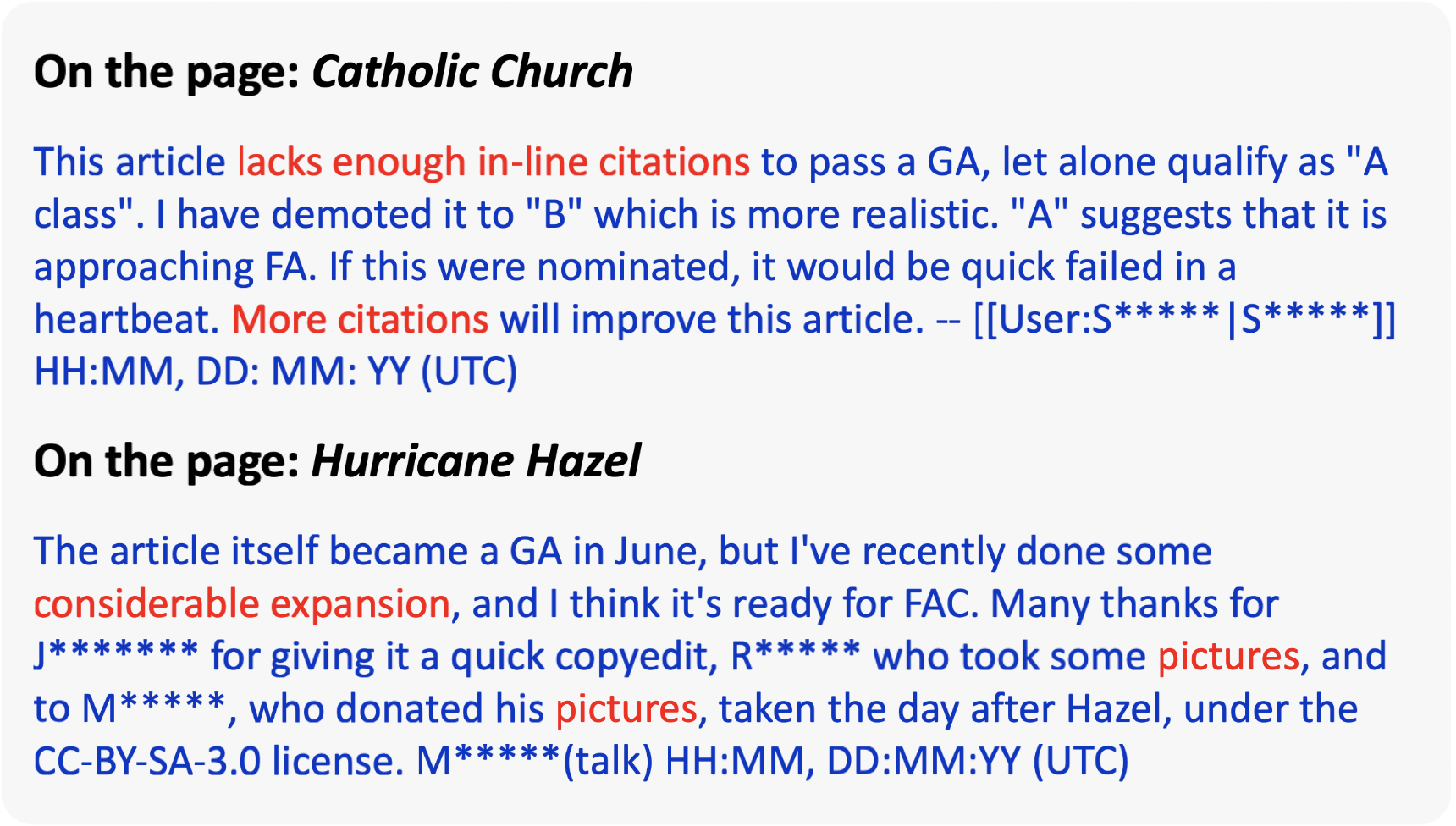}
    \caption{A snapshot of talk page conversations from a couple of randomly chosen pages that involve discussion for quality change of those pages.}
    \label{fig:qualitative_examples}
\end{figure}

\noindent\textbf{Success of our approach}: We revisited a couple of articles and tried to compare the detected change points of quality switch predicted by our models with the ground truth. As noted in the previous sections, the talk pages contain heightened discussions among the editors before any quality change. Editors put forward their opinions, either converging to an agreement or landing at a conflict. We discuss two interesting cases here as depicted in the Figure \ref{fig:qualitative_examples}. In the first case of \textit{Catholic Church} article, the editor is found to oppose the current higher quality (AGA) and demoted the quality to B class, which is a prominent example of opinion conflict. In the other example, \textit{Hurricane Hazel}, editors had reached a consensus to promote the current quality of the article to the FA class. In both the cases, we include the timestamp mentioned in the talk pages as the change of quality and execute our methods. We observe that our proposed approach detect the change points accurately within the accepted margin of error \textit{M}. Further, the detection of quality switch coincided with the actual change points, which we found was primarily triggered by an abrupt change in the content of the article. 

\subsection{Design implications}
\blu{Change point detection (CPD) task in time series analysis has achieved significant importance in numerous fields such as climatology~\cite{verbesselt2010detecting}, network traffic data analysis~\cite{lung2012distributed} etc. Similarly, the CPD algorithms are well studied in CSCW/peer-production communities. In the domain of software and services, authors in their work~\cite{daly2020use} have described implementation of change point detection system on the basis of E-Divisive means algorithm to identify the commits responsible for performance regressions. In the setup of continuous integration (CI) system of software products, performance tests, i.e., a large number of benchmark tests are run periodically (e.g., in every 2 hours or every 24 hours) and try to visualize the results (performance trend graph) as time series over time. Inclusion of CPD algorithms into CI system helps the authors finding changes in performance by dropping false positive rate as well as catching smaller regressions in a timely manner. In another challenging task of detecting changes in online conversation~\cite{goutte2018eurogames16}, authors have 
applied change point detection algorithms to detect salient changes within an existing event or story line. The authors collected a stream of messages from Twitter related to a specific topic and thus the dataset of 16 sports events are semantically converted into one or several time series, describing the sentiment or topic of the conversation. Then they applied change point detection algorithms to identify the locations of significant changes from the linguistically motivated signals of online conversation. In general CPD algorithms can be used to track topic change or even topical sentiment change in online peer-production systems (e.g., collection of new articles etc.).} In our work, since our features are very lightweight and very easily interpretable it would open up new opportunities for content governance by the Wikipedia community who can tinker our model architecture, the feature space, the hyperparameters etc. to suit their purposes. These are the well known components of \textit{dependency injection}\footnote{\url{https://en.wikipedia.org/wiki/Dependency_injection}} which allows for injection of any new feature(s) in the pipeline and \textit{threshold optimisation} which allows for tuning all the different thresholds in the pipeline (e.g., $M$ in our CPD algorithms). Likewise ORES \cite{halfaker2019ores}, we would also open source our data and the models to promote further development and proliferation of the alert system as a socio-infrastructural service. This therefore means that the Wikipedia communities do not need our explicit intervention in the governance of the pipeline and the alert system thereby, developed by us. 

In fact, the online world is increasingly getting crammed with for-profit user generated content. Wikipedia stands out as a rare exception in this race. However, one of the major bottlenecks is scaling such a platform, that allows anyone to edit, for the good. Hence efficient machine learning algorithms to facilitate content moderation is a dire need of the day. This has led to the new initiative called ``participatory machine learning''~\citep{paml} that imagines machine learning models to be just as provisional and open to criticism, revision etc. as are the contents of any Wikipedia article.  In fact, ORES was introduced by the Wikimedia foundation to achieve this purpose. Our work being a socio-technical and CSCW-oriented approach can be directed to the same lines in future and is expected to open up new collaborative practices by generating timely alerts for the editors. We believe that such collaborative practices should further enable faster resolution of disputes in the article content that is presently done using RfCs (Requests for Comments)~\cite{im2018deliberation}.

\subsection{Quality change: norm or exception?}
\blu{Is it surprising that more than 50\% of the articles don't experience any quality change over their entire lifespan? It would therefore not be surprising to see a long-tail on Wikipedia, where most articles do not change their quality frequently, while a few articles do change their quality regularly. So how serious is this problem? Going by the observations in Figures~\ref{fig:user_views},~\ref{fig:editions} and~\ref{fig:collabs} it becomes apparent that there is preferential treatment of certain pages. Typically, pages that have large page views, many editions and a large number of collaborators experience more frequent quality changes. Indeed, therefore a long tail behaviour is expected. However, we would like to point out that this is in contrast to Wikipedia's neutral point of view (NPOV) guidelines\footnote{\url{https://en.wikipedia.org/wiki/Wikipedia:Neutral_point_of_view}}. For instance, there is a long standing objection that Wikipedia is dominated by Anglo-American content\footnote{\url{https://en.wikipedia.org/wiki/Wikipedia:Neutral_point_of_view/FAQ\#Other_objections}}. The website itself acknowledges this to be ``an ongoing problem that should be corrected by active collaboration between Anglo-Americans and people from other countries". In fact, a new Wikiproject\footnote{\url{https://en.wikipedia.org/wiki/Wikipedia:WikiProject_Countering_systemic_bias}} to counter such systemic biases which are possibly an outcome of the well-known disparate impact~\cite{solon16}, has been set up by Wikipedia. Our work echoes the same concerns albeit in a rigorous quantitative fashion.}

\subsection{Limitations}

One of the limitations of our methods like many other machine learning models is the lack of explanation of the results predicted with the increased dimension of features. However, since the entire framework can be made participatory, there can always be options created for the Wikipedia community using our alert system to refute, support, discuss and critique the predictions made by the system. We plan to develop a framework in the future that would allow the community to report mistakes made by our model in a structured and streamlined fashion. In fact, a database of contextual mistakes made by our model can be maintained and the community could query the database to understand the reasons for misclassifications better. This in the long run is expected to help in the genesis of fully explainable model.

\section{Conclusion}
In this paper, we have inspected the life cycle of article quality over a massive dataset of $30k$ articles and noticed varying dynamical patterns of quality moderation. One of the most important findings of our work is that while more than 50\% articles do not experience any quality change over their entire lifespan, there are some articles that undergo quality shifts multiple times and that too within very short spans of time. We also observed possibilities of quality switch wars apparent from the rapid cyclic switches on various Wikipedia articles. As a second objective we modeled the detection of quality changes as a change point detection problem. We formulated a series of very intuitive features and fed them to standard change point detection algorithms. We obtained decent scores in terms of different evaluation metrics. One important thing to notice here is that the content based features are themselves sufficient to trigger the best results. This modeling, we believe shall pave the way to the design of a full-fledged alert system for the Wikipedia community so that they can take a more informed, proactive and timely decision on quality changes related to an article. Our belief is further strengthened by the following post-facto analysis of the results. 

\if{0}\textcolor{red}{
The online world is increasingly getting crammed with for-profit user generated content. Wikipedia stands out as a rare exception in this race. However, one of the major bottlenecks is scaling such a platform, that allows anyone to edit, for the good. Hence efficient machine learning algorithms to facilitate content moderation is a dire need of the day. This has led to the new initiative called ``participatory machine learning''\citep{paml} that imagines machine learning models to be just as provisional and open to criticism, revision etc. as are the contents of any Wikipedia article.  In fact, ORES was introduced by the Wikimedia foundation to achieve this purpose. Our work being a socio-technical and CSCW-oriented approach can be directed to the same lines in future and is expected to open up new collaborative practices by generating timely alerts for the editors. We believe that such collaborative practices should further enable faster resolution of disputes in the article content that is presently done using RfCs (Requests for Comments)}~\cite{im2018deliberation}.\fi

As an immediate future step we plan to extend the work to a multilingual setting. We would also like to employ more complex change point detection models (the current models are chosen based on their easy interpretability). We would finally wish to deploy a realistic early quality change monitoring system on the Wikipedia platform and observe the effectiveness of the same in quality moderation.  

\if{0}\noindent
Further we correlated the dynamics of qualities of articles over their entire lifespan with the comprehensive approach of time series. We employ the state-of-the-art approaches of change point detection in early prediction of qualities. Our simple handcrafted feature driven approach included different factors that play a key role in assigning qualities. Although our methodology is completely unsupervised, it secured the best score of covering 0.6 with the combination of edit pattern based features along with the editor and content based features.\fi

\bibliographystyle{ACM-Reference-Format}
\bibliography{ref}

\begin{appendix}
\section{Appendix}

\subsection{Turn around time}\label{appendix:box}
\begin{figure}[h!]%
    \centering
        \subfloat[\centering \label{fig:box_promotion} Only promotion in quality. changes]{{\includegraphics[height=3cm,width=0.45\columnwidth]{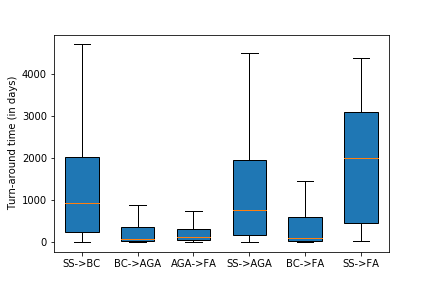}}}%
    \qquad
    \subfloat[\label{fig:box_demotion}\centering Only demotion in quality changes.
    ]{{\includegraphics[height=3cm,width=0.45\columnwidth]{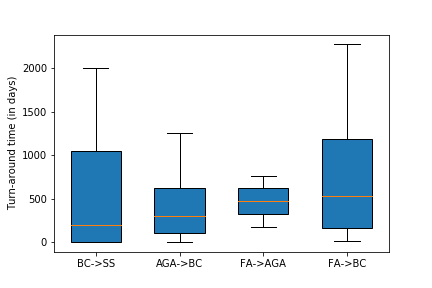}}}%
    \qquad
    \subfloat[\label{fig:box_nochange}\centering No change in quality. ]{{\includegraphics[height=3cm,width=0.45\columnwidth]{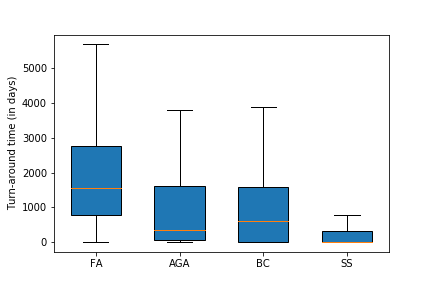}}}
    \caption{The box-plots showing turn-around time distribution for three categories -- (a) only promotion, (b) only demotion, (c) no change in quality.}
    \label{fig:boxplot}
\end{figure}

Here we report the box-plots of the turn-around time distributions using the for three major categories of quality change pattern - (a) only promotion, (b) only demotion, (c) no change in quality (please see Figure~\ref{fig:boxplot}). It can be observed that turn around time is not consistent across the articles for all types of quality changes, specifically in case of longer hops (e.g., promotion from SS to FA or demotion from FA to BC). The box plots show the evidence of delayed quality assessment which is natural in case of long jumps of quality changes. For all the cases of no change in quality, the statistics looks similar except the SS. The pages belonging to the SS category were always assessed immediately after their creation. 
 
\subsection{Temporal patterns of the different features}\label{appendix:A}

\begin{figure}[h!]
\subfloat{
\includegraphics[width=0.3\textwidth]{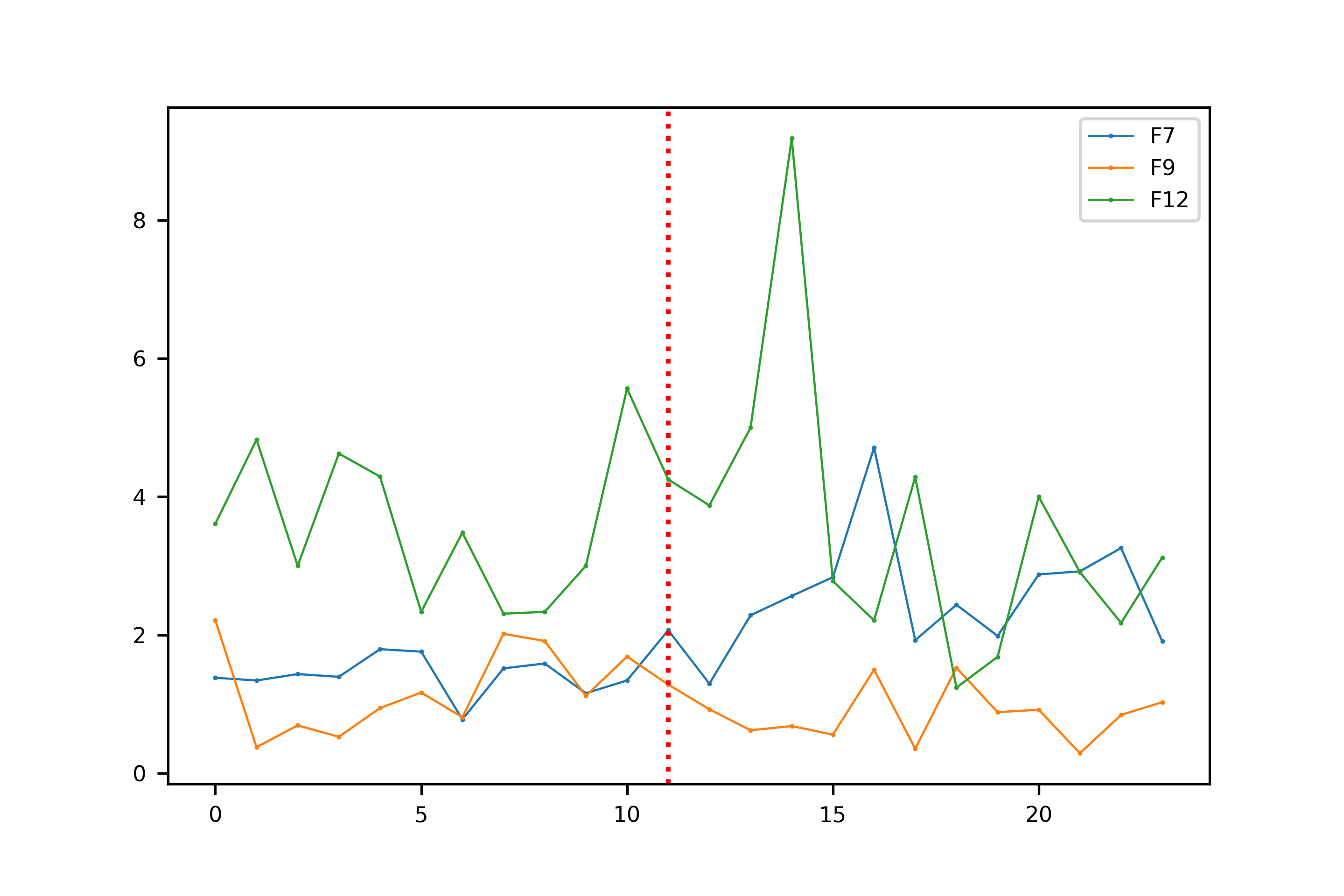}}\qquad    
\subfloat{
\includegraphics[width=0.3\textwidth]{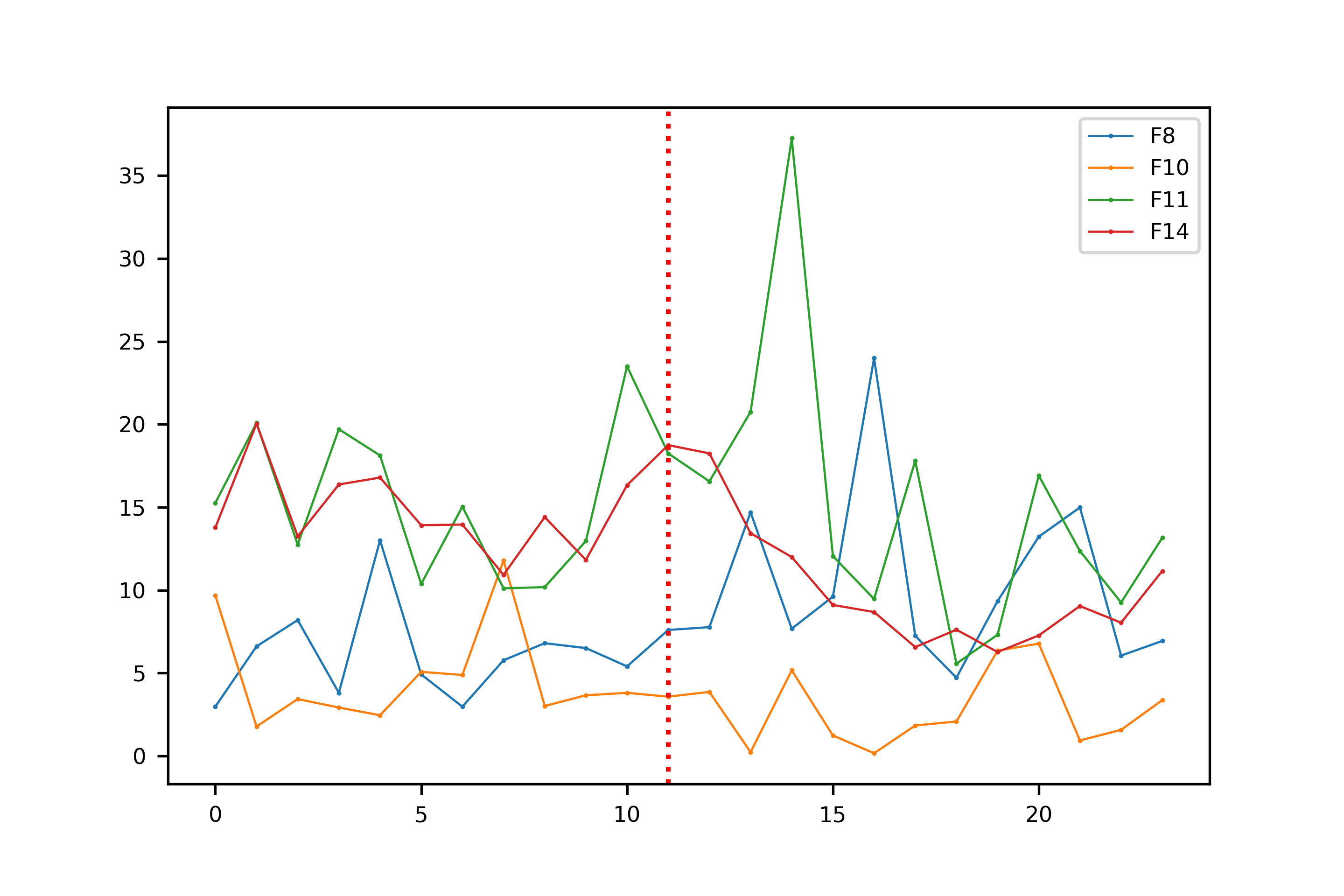}}\qquad    
\subfloat{
\includegraphics[width=0.3\textwidth]{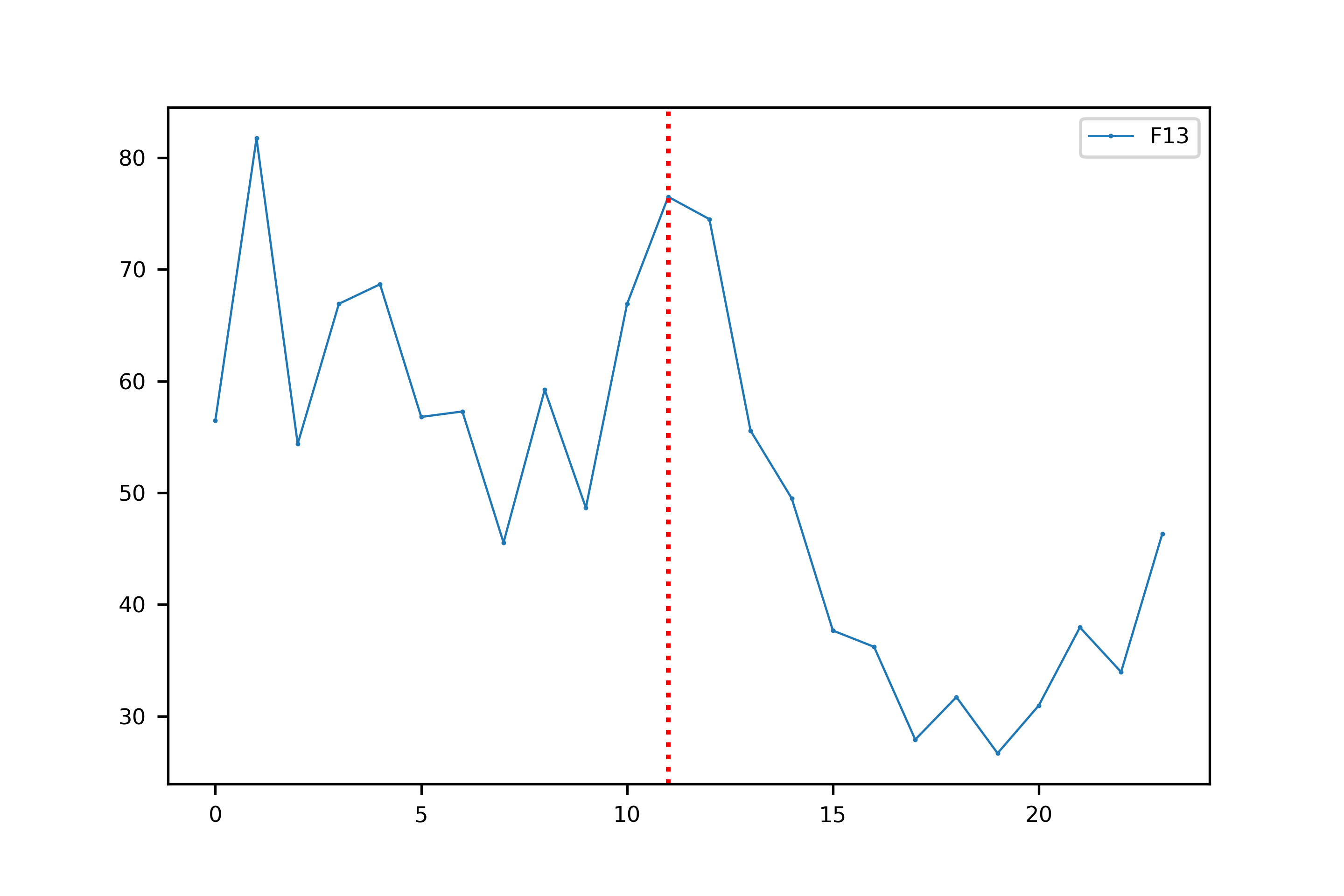}}   \qquad 
\subfloat{
\includegraphics[width=0.3\textwidth]{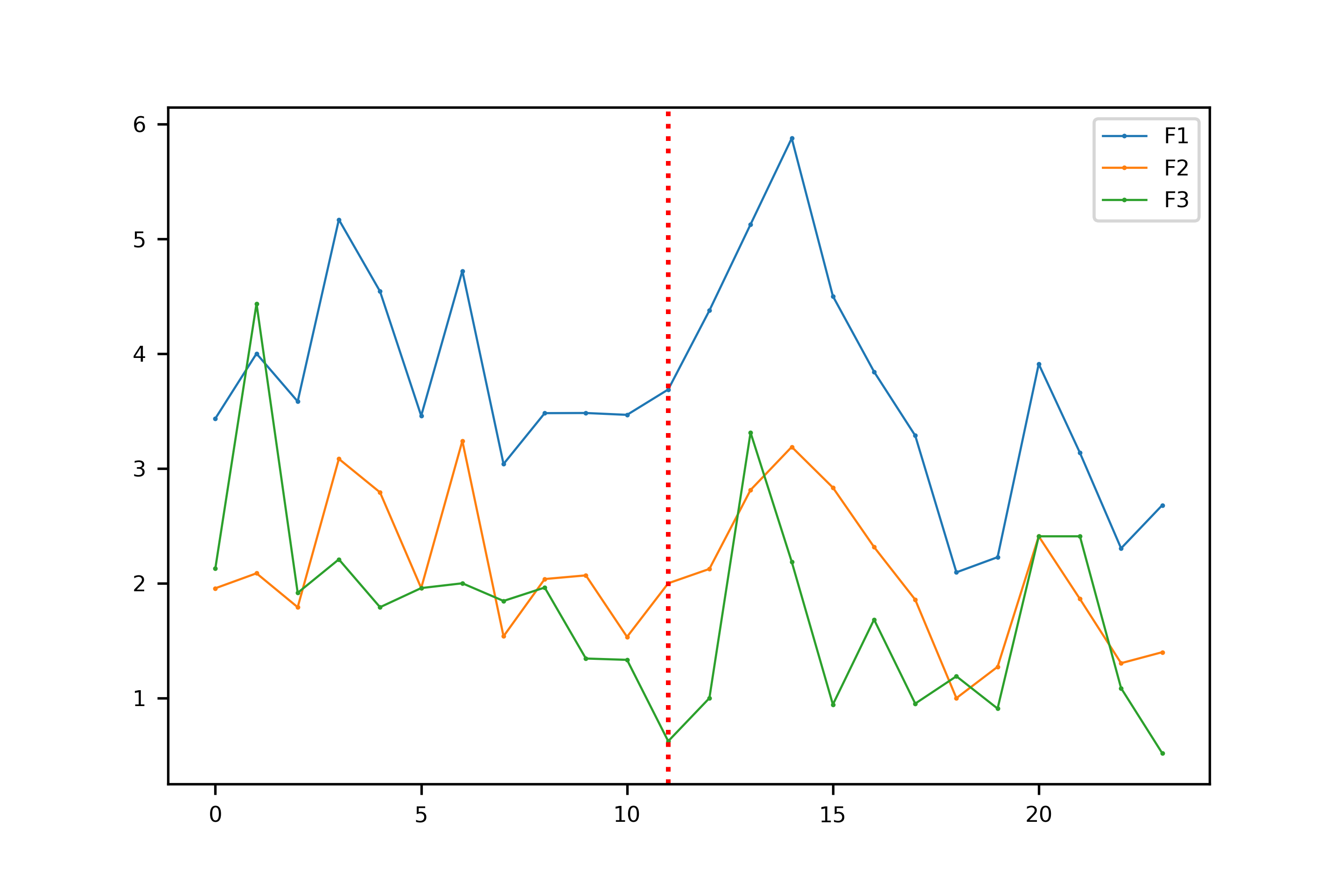}}  \qquad  
\subfloat{
\includegraphics[width=0.3\textwidth]{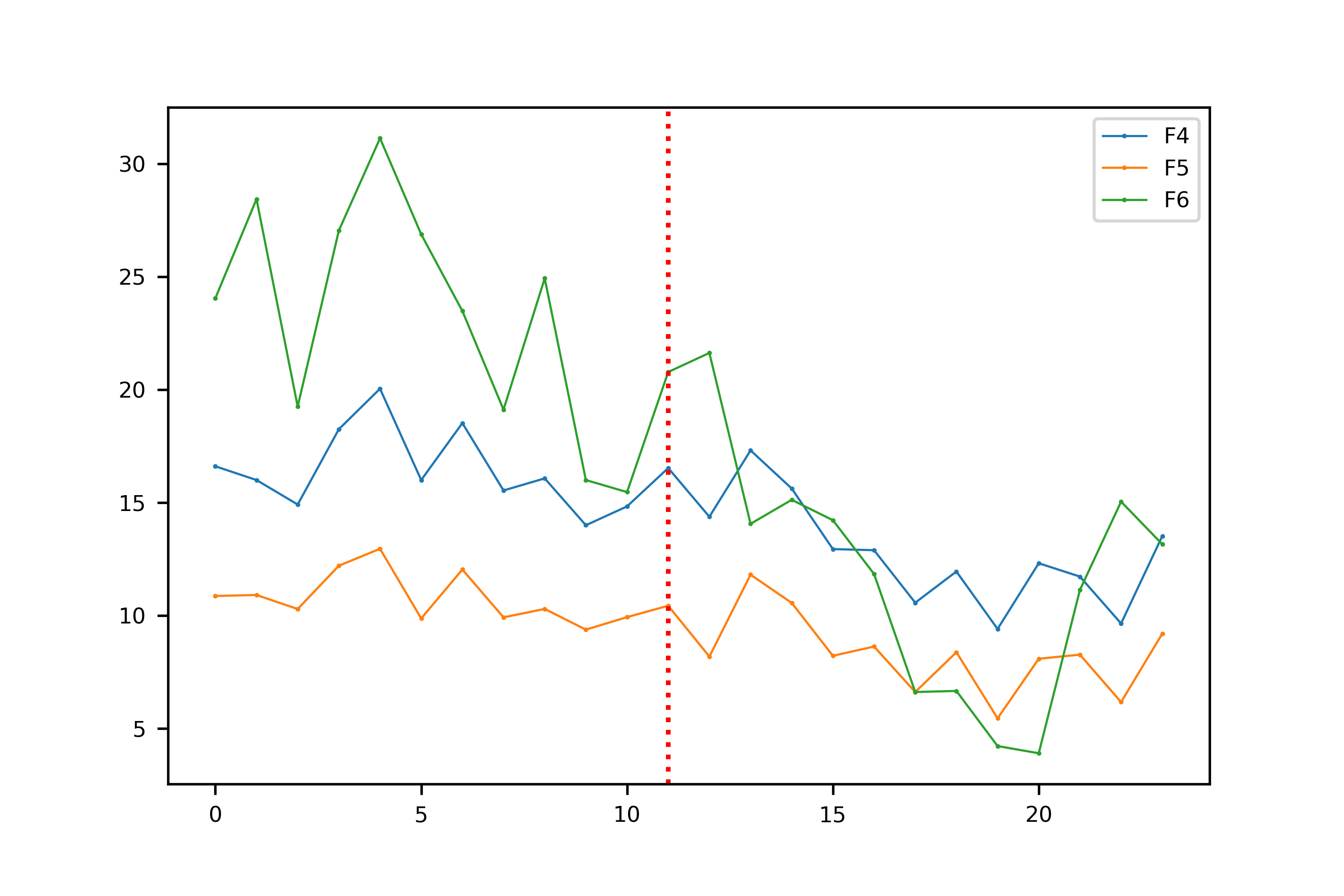}}
\caption{Plots showing temporal trends in contribution based and activity based features in case of 54 sample articles.}
\label{fig:temporal_editor_activity}
\end{figure}

\begin{figure}[h!]
\subfloat{
\includegraphics[width=0.28\textwidth]{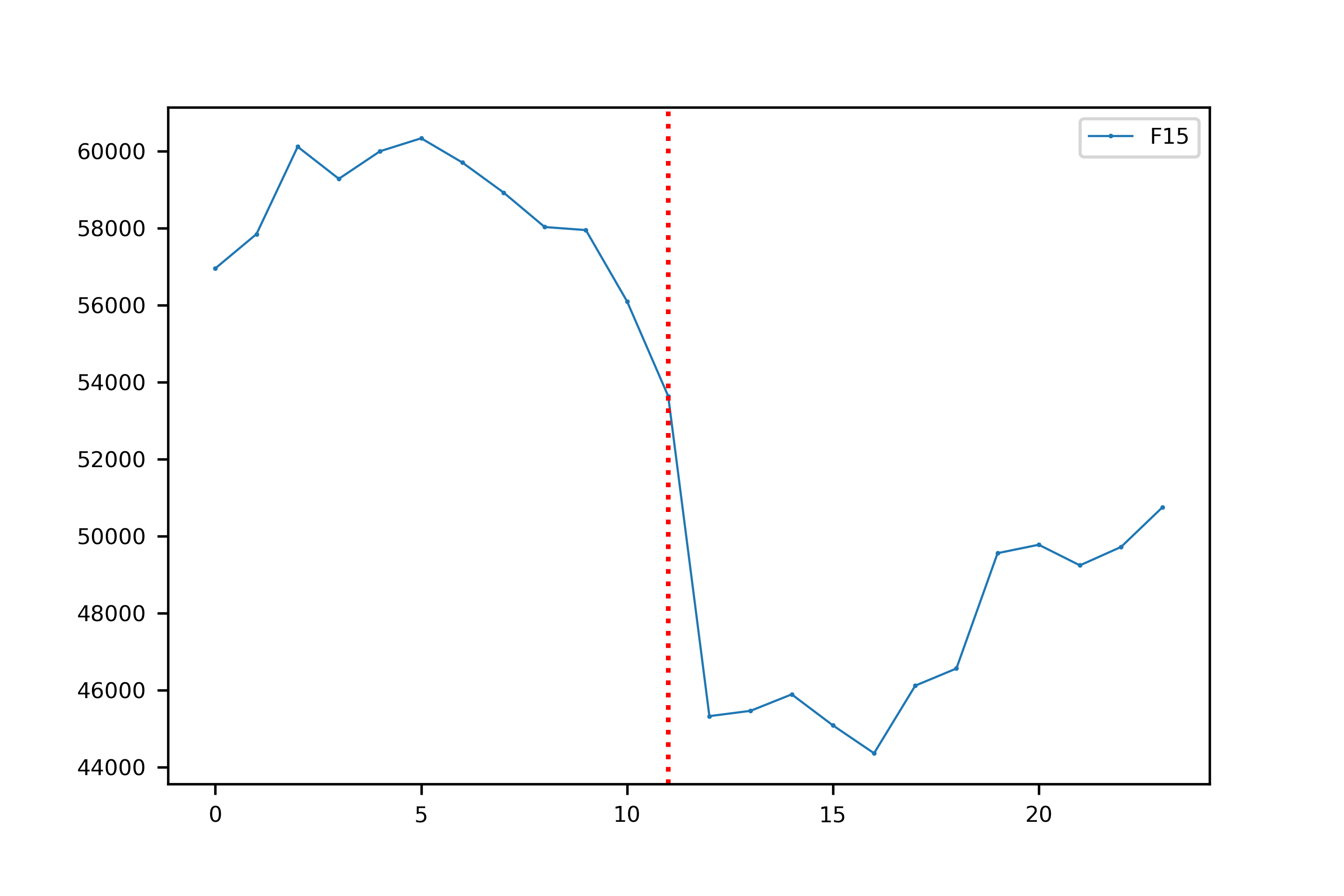}} \qquad
\subfloat{
\includegraphics[width=0.28\textwidth]{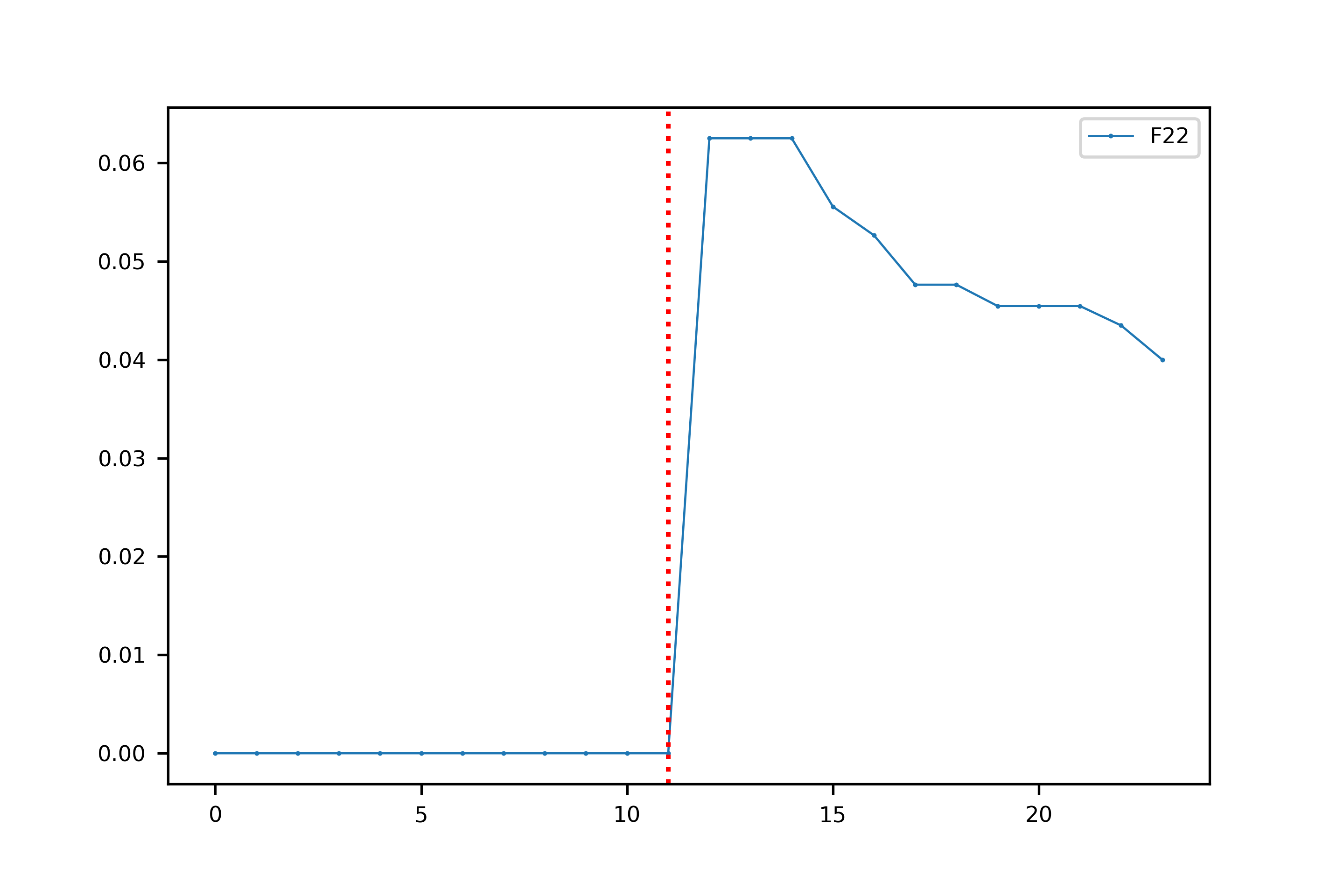}} \qquad
\subfloat{
\includegraphics[width=0.28\textwidth]{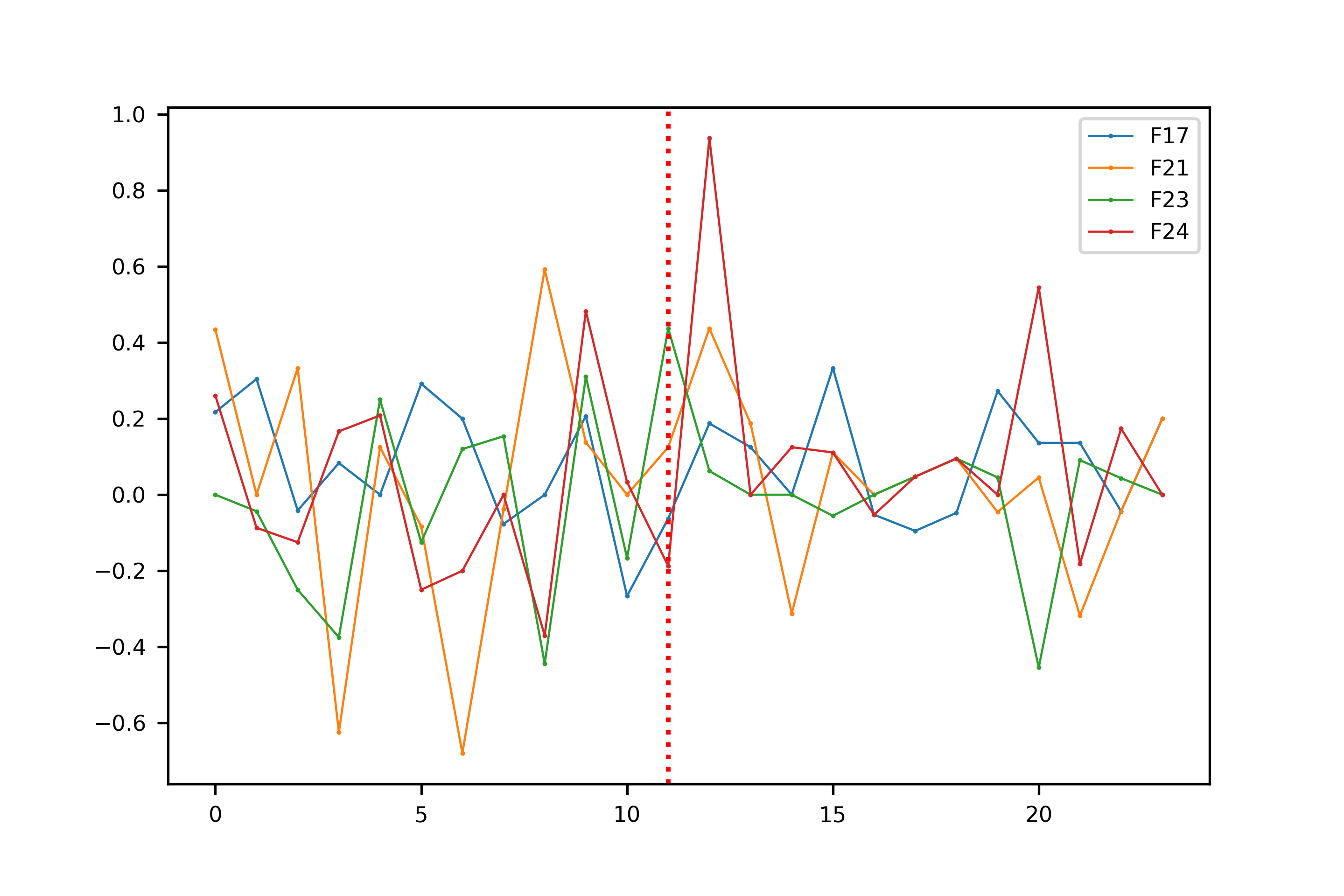}} \qquad
\subfloat{
\includegraphics[width=0.28\textwidth]{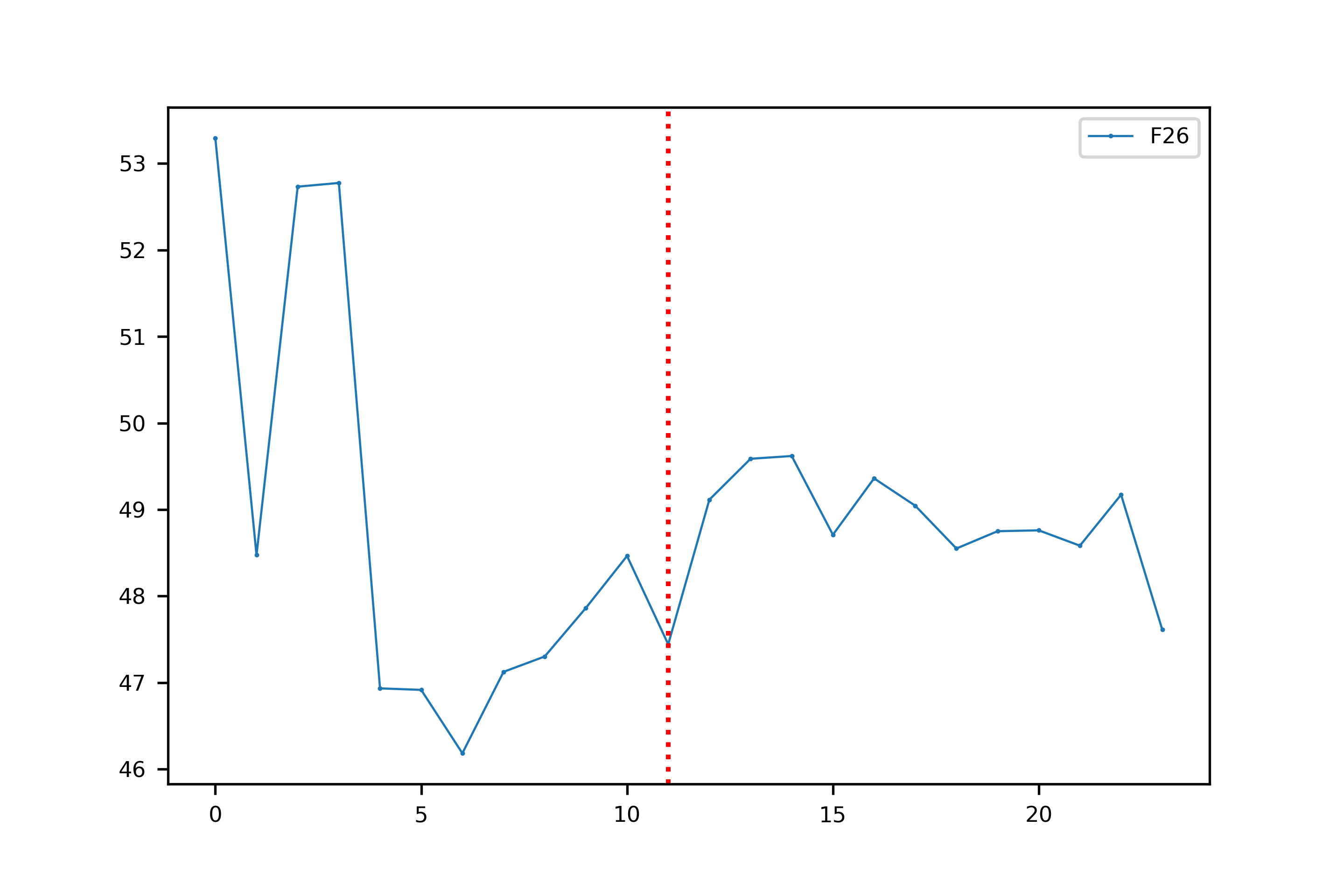}}    \qquad
 \subfloat{
\includegraphics[width=0.28\textwidth]{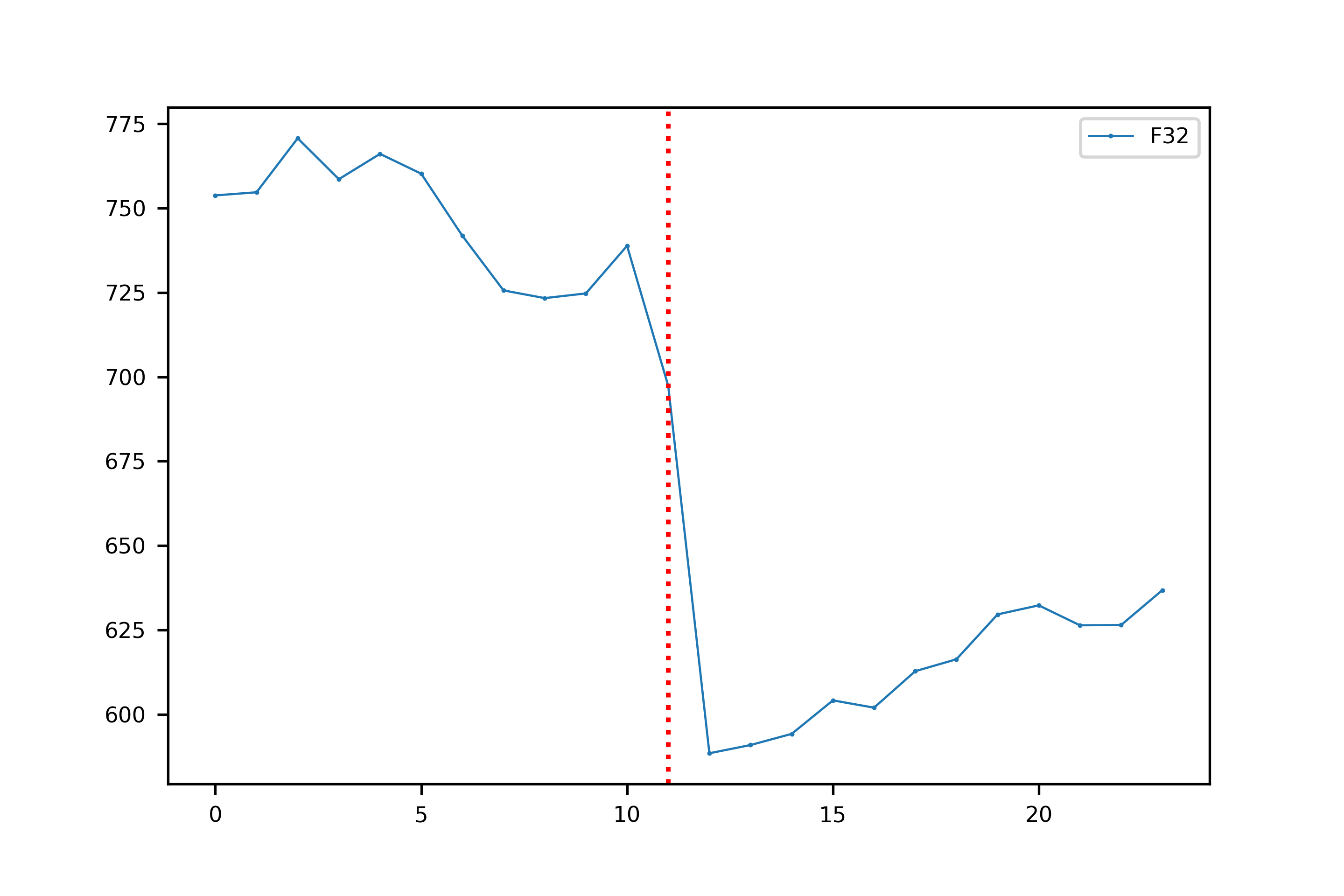}} \qquad
\subfloat{
\includegraphics[width=0.28\textwidth]{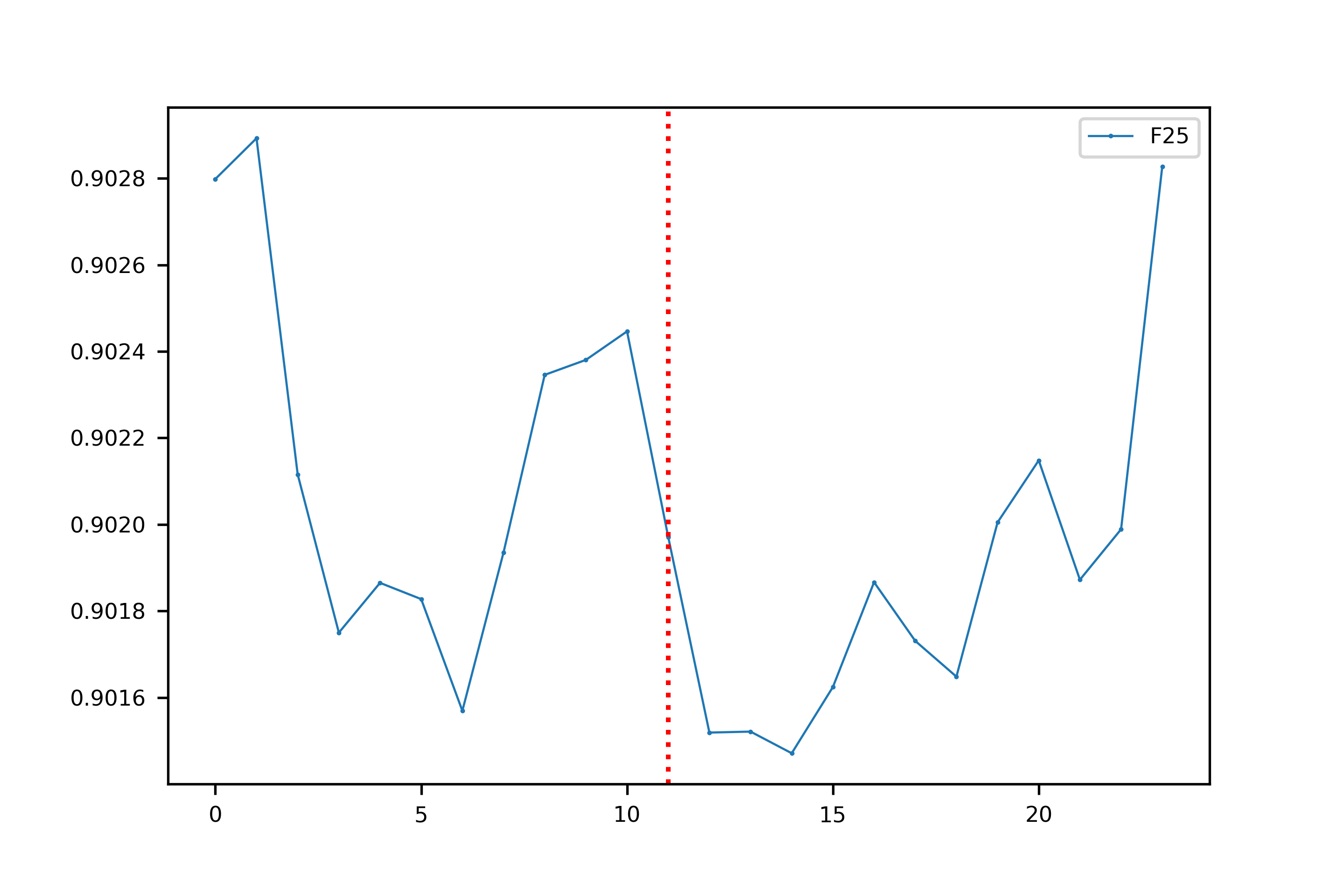}}    \qquad
\subfloat{
\includegraphics[width=0.28\textwidth]{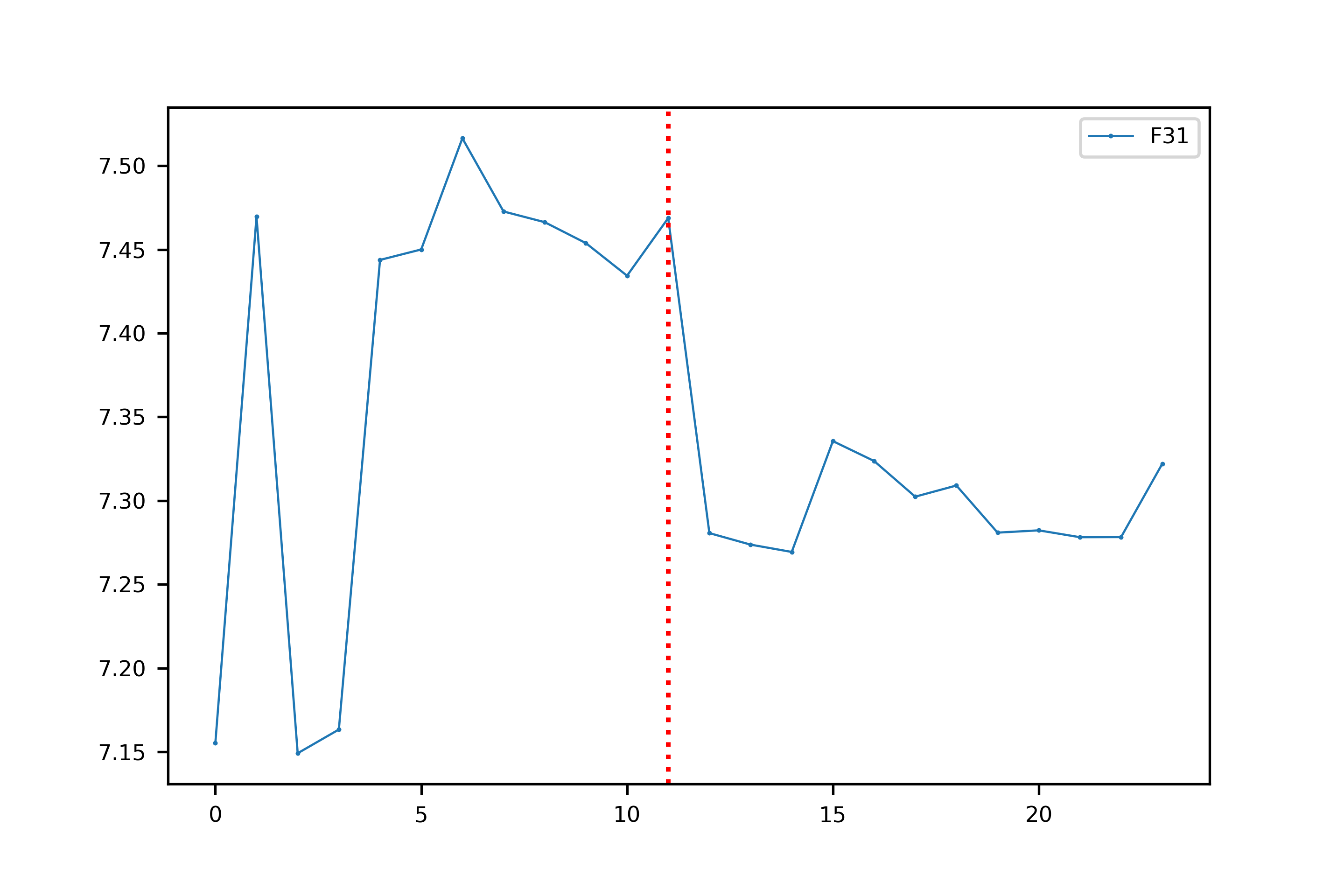}} \qquad
\subfloat{
\includegraphics[width=0.28\textwidth]{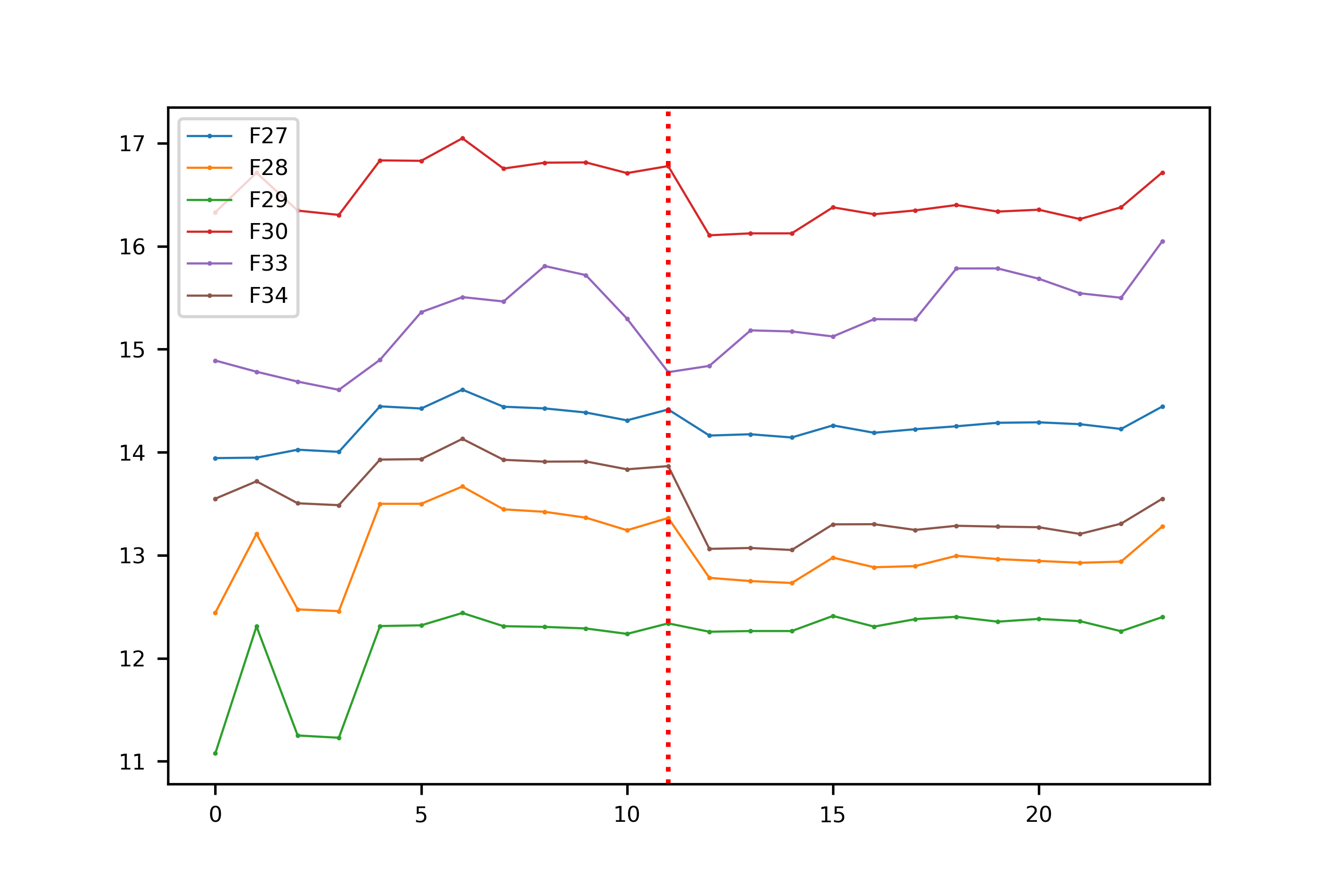}}    \qquad
\subfloat{
\includegraphics[width=0.28\textwidth]{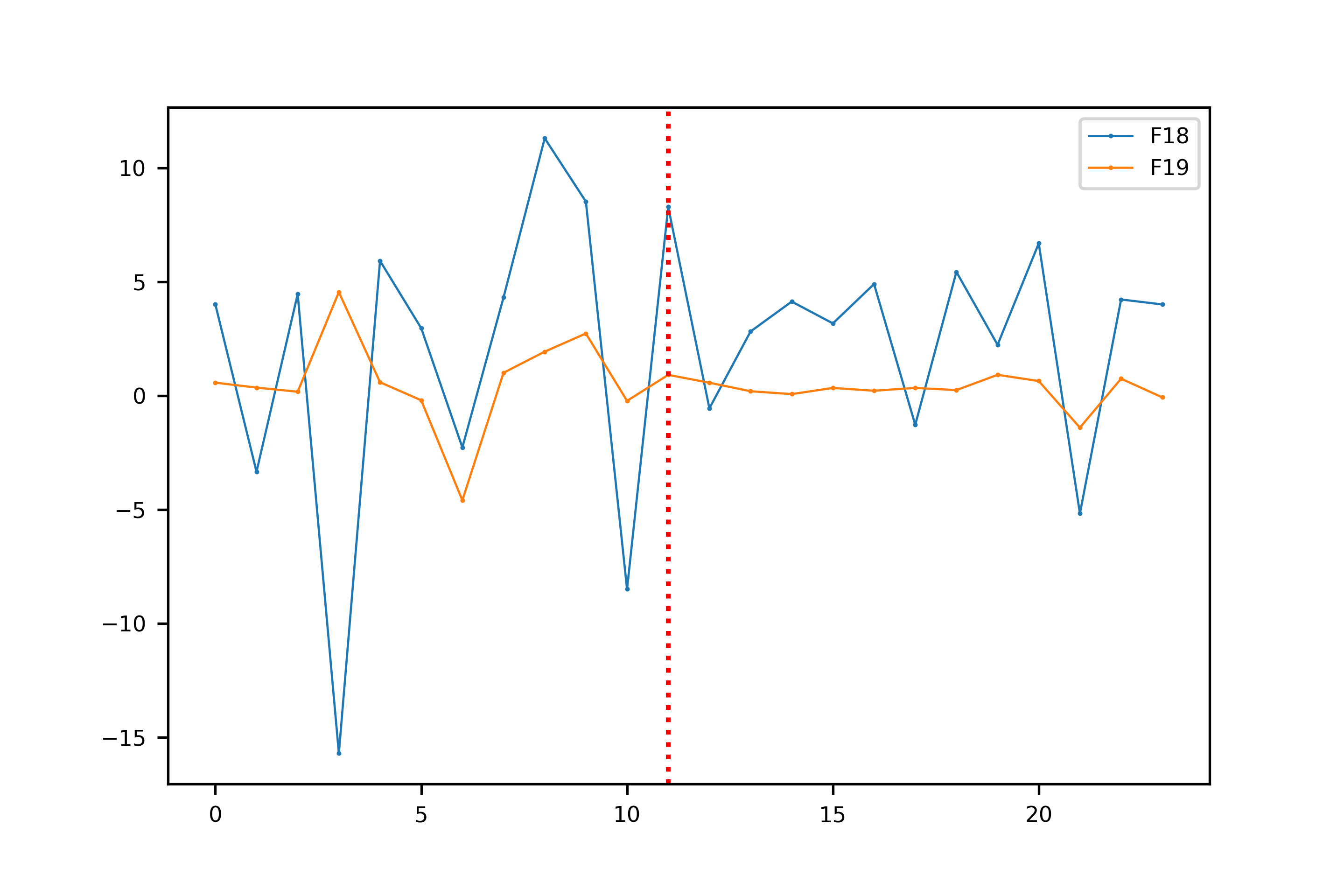}} \qquad
\subfloat{
\includegraphics[width=0.28\textwidth]{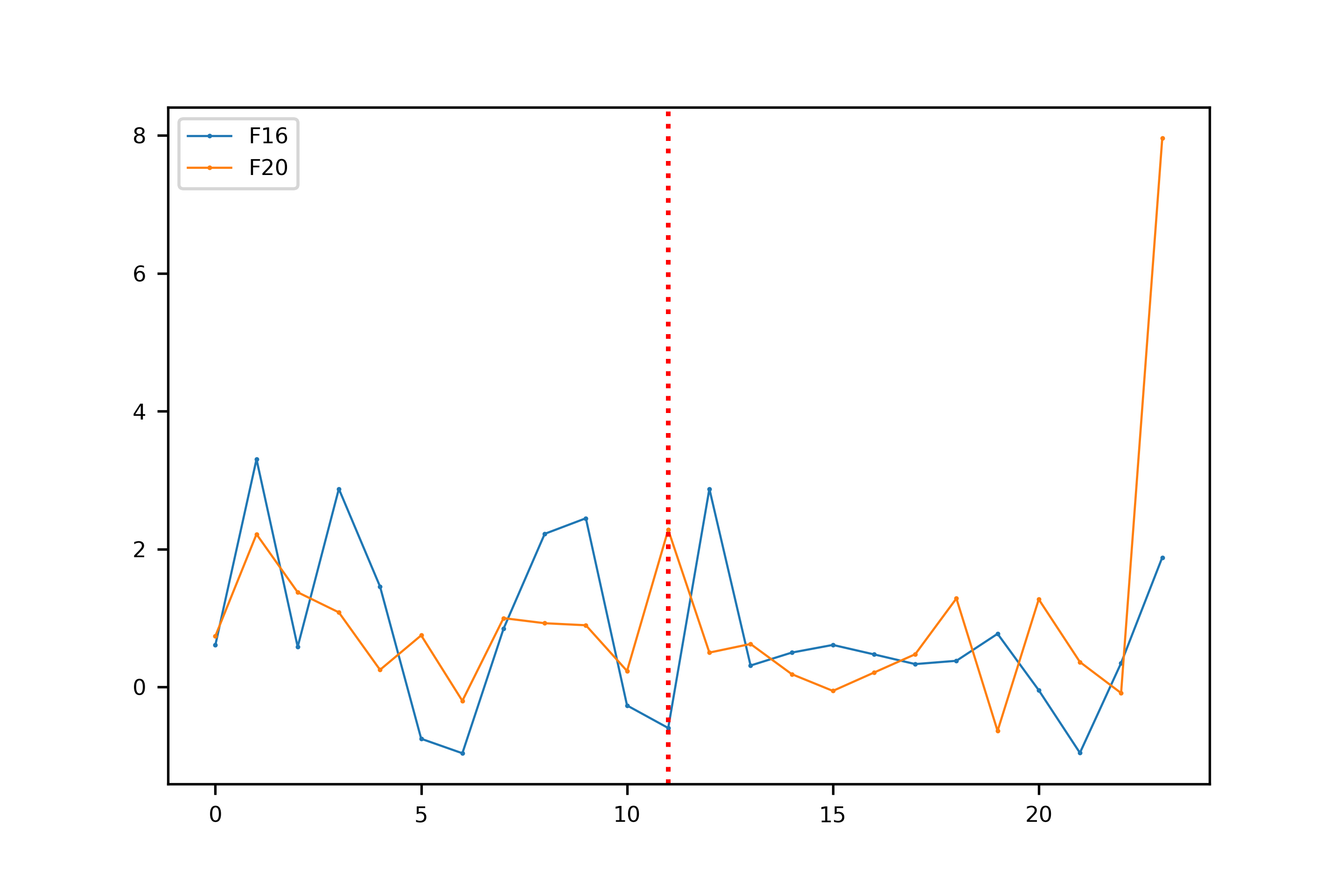}}    \qquad
 \caption{Plots showing temporal patterns observed in content based features for the sample set of 54 articles.}
\label{fig:tempral_content}
\end{figure}

Here we report the temporal trends of all the 34 features for the 54 pages that we chose for our experiments. According to our hypothesis, we observed significant change in temporal trend at the junction of quality shift (i.e, from FA to lower quality) for all the features in the data sample of 54 pages.  (please see Figures~\ref{fig:temporal_editor_activity} and~\ref{fig:tempral_content}).

\end{appendix}

\end{document}